\documentclass[galaxies,article,accept,moreauthors,pdftex]{Definitions/mdpi}

\firstpage{1} 
\pubvolume{1}
\issuenum{1}
\articlenumber{0}
\usepackage{amsmath,amssymb}
\usepackage{amsmath}
\DeclareMathOperator{\arcsec}{"}
\pubyear{2021}
\copyrightyear{2020}
\datereceived{} 
\dateaccepted{} 
\datepublished{} 
\hreflink{https://doi.org/} 
\pdfoutput=1

\Title{The {\it e}MERLIN and EVN view of FR\,0 radio galaxies}

\TitleCitation{The EVN and {\it e}MERLIN view of FR\,0 radio galaxies}

\Author{Ranieri D. Baldi $^{1,2,*}$\orcidA{}, Gabriele Giovannini $^{1,3}$ and Alessandro Capetti$^{4}$}

\AuthorNames{Ranieri D. Baldi, Gabriele Giovannini and Alessandro Capetti}

\AuthorCitation{Baldi, R.~D.; Giovannini, G.; Capetti, A.}

\address{%
$^{1}$ \quad INAF - Istituto di Radio Astronomia, via P. Gobetti 101, I-40129, Bologna, Italy; ranieri.baldi@inaf.it\\
$^{2}$ \quad School of Physics and Astronomy, University of Southampton, Southampton, SO17 1BJ, UK\\
$^{3}$ \quad Dipartimento di Fisica e Astronomia, Universit\`a di Bologna,  via Gobetti 93/2, 40129 Bologna, Italy\\
$^{4}$ \quad INAF - Osservatorio Astrofisico di
Torino, Strada Osservatorio 20, I-10025 Pino Torinese, Italy}

\corres{Correspondence: ranieri.baldi@inaf.it}

\abstract{We present the results from high-resolution observations carried out with the {\it e}MERLIN UK-array and the European VLBI network (EVN) for a sample of 15 FR~0s, i.e. compact core-dominated radio sources associated with  nearby early-type galaxies (ETGs) which represent the bulk of the
local radio galaxy population. The 5-GHz {\it e}MERLIN observations available for 5 objects exhibit sub-mJy core components, and reveal pc-scale twin jets for 4 out of 5 FR~0s once the {\it e}MERLIN and JVLA archival visibilities data are combined. The 1.66-GHz EVN observations available for 10 FR~0s display one- and two-sided jetted morphologies and compact cores. The pc-scale core emission contributes, on average, to about one tenth of the total extended radio emission, although we note a increasing core contribution for flat/inverted-spectrum sources. We found an unprecedented linear correlation between the pc-scale core luminosity ($\sim$ 10$^{21.3}$-10$^{23.6}$ W Hz$^{-1}$) and [O~III] line luminosity, generally considered as proxy of the accretion power, for a large sample of LINER-type radio-loud low-luminosity active nuclei, all hosted in massive ETGs, which include FR~0s and FR~Is. This result represents further evidence of a common jet-disc coupling in FR~0s and FR~Is, despite they differ in kpc-scale radio structure. For our objects and for other FR~0 samples reported in the literature, we estimate the jet  brightness sidedness ratios, which typically range between 1 and 3. This parameter roughly gauges the jet bulk Lorentz factor $\Gamma$, which turns out to range between 1 and 2.5 for most of the sample. This corroborates the scenario that FR~0s are characterized by mildly-relativistic jets, possibly as a result of lower spinning black holes (BHs) than the highly-spinning BHs of relativistic-jetted radio galaxies, FR~Is.}

\keyword{radio continuum: galaxies; galaxies: active}

\begin{document}

\section{Introduction}

The presence of accreting supermassive black holes (BHs) (Active Galactic Nuclei, AGN) at the centre of local massive early-type galaxies (ETGs) has been largely reported by large-area surveys (e.g. \cite{ferrarese00,schawinski07,boroson11,nyland16,werner19,grossova21}) and  supported by theoretical studies (e.g. \cite{croton06,sijacki07,fabian12}). The most massive ETGs appear to be typically associated with radio-loud AGN (RL AGN or radio galaxies, RG, in general) able to launch the most powerful jets in the Universe. In opposition to the latter which have typically activity period of 10$^{7}$-10$^{8}$ yr, low-power ($\lesssim 10^{23}$ W Hz$^{-1}$) RGs appear to be steadily active although at low regime \cite{sabater19}. From the most powerful to the weakest, RGs  encompass a large range of jet properties, such as luminosities, morphologies, duty cycles and speeds, but all sharing a single type of evolved host \cite{heckman14,morganti17}. This capability of massive ETGs to be frequently associated with radio AGN has been recently object of several  observational and theoretical studies and contextualised in the framework of AGN feedback, since RGs, even at low powers, can continuously inject energy in the central kpc of galaxies (e.g. \cite{best05a,fabian12,hardcastle19,hardcastle20,webster21}).

The vast majority of local massive ETGs are characterised by low-luminosity compact flat-spectrum radio sources (e.g. \cite{rogstad69,heeschen70,ekers73,kellermann81,sadler84,wrobel91,slee94,giroletti05,hardcastle19}). Apart from Seyferts and star-forming galaxies, where several radio-emitting thermal and non-thermal mechanisms can co-exist \cite{panessa19}, the majority of the most prominent flat-spectrum nuclear radio sources in nearby galaxies are found in galaxies with LINER
(Low-Ionization Nuclear Emission line Regions \cite{heckman80}) nuclear spectra \cite{nagar00}. They are 1) about two-third of the all local active galaxies \citep{nagar05,saikia18}; 2) characterized by low-ionization optical forbidden lines \cite{ho08} whose source of ionization has been debated for decades (low-luminosity version of bright AGN \cite{halpern83}, shocks \cite{dopita95}, or stellar photoionization \citep{binette94,capetti11}); and 3) powered by low-accreting radiatively inefficient discs (RIAF, \cite{narayan95}) in case of hosting an active BH \cite{falcke04}. The accretion-powered LINERs tend to host compact radio louder cores \cite{cohen69,falcke00,filho02,maoz07}, as the BH mass (or galaxy mass) increases (e.g. \cite{laor00,best05a,mauch07,baldi21b}). In fact, as LINERs (or in RL regime, Low Excitation Radio Galaxies\footnote{In the optical taxonomy of RL AGN, RL LINERs are typically named as Low Excitation Radio Galaxies (LERGs), while Seyferts as High Excitation Radio Galaxies \cite{hine79,buttiglione10}. LINER and LERG are, thus, equivalent optical classifications. Local FR~Is are almost totally classified as LINERs, whereas FR~IIs are associated with either LINER or Seyfert-like nuclear spectra. For example, In the Third Cambridge (3C) catalogue of RGs \cite{bennett62}, FR~Is are mostly all LERGs, with only a few possible candidates of FR~I HERGs, e.g. 3C~120.}) are usually radio-louder than the other optical classes and similar to low-power RGs \cite{nagar02,capetti06,kharb12a}, a model of 
synchrotron self-absorbed base of a low-power jet coupled with underluminous RIAF disc (typically in the form of an advection-dominated accretion flow, ADAF \cite{narayan94,narayan98,narayan08}), analogous to FR~I disc-jet interpretation (e.g. \cite{balmaverde06b,hardcastle09}), has been used to describe their accretion-ejection mode. Low-luminosity LINERs have been generally interpreted as the scaled-down version of  the cores of powerful RL AGN, showing a very high brightness temperature and a flat to inverted radio spectrum that extends up to sub-mm wavelengths \cite{ho99,nemmen14}.

Recently, compact radio sources associated with massive ETGs with LINER nuclear spectra have been  named FR~0 radio galaxies \cite{ghisellini11,sadler14} in contrast with the other Fanaroff-Riley classes \cite{fanaroff74} because of lack of substantial extended emission \cite{baldi09,baldi10,sadler16,baldi16}. FR~0s are also characterised by  a core dominance, which is about 30 times higher than that of FR~Is \cite{baldi15,whittam20}, and a poorer Mpc-scale environment (living in groups containing less than $\sim$15 galaxies) \cite{capetti20b} than FR~Is. FR~0s represent the most abundant population of RGs, being at least five times more numerous than FR~Is in the local Universe (redshift z $<$ 0.05) \cite{baldi18a}. Most of FR~0s have been studied at arcsec resolution with NVSS and FIRST datasets with a compact unresolved morphology. Concerning their radio-band spectral properties, their SED are generally flat (or flattening at higher frequencies) from  hundreds of MHz (arcsec resolution with TGSS and LOFAR, \cite{capetti19,capetti20a}) to a few GHz (sub-arcsec resolution with JVLA) and recently even to higher radio frequencies ($\sim$15-22 GHz at arcmin angular resolution with compact array or single dish \cite{mikhailov21,baldi22}). A multi-frequency JVLA follow-up observations of 25 FR~0s at sub-arcsec resolution reveal the presence of kpc-scale jets for about one fourth of the sample \cite{baldi15,baldi19}, the first evidence of the presence of clear extended jets in FR~0s. 
In terms of nuclear and host properties, FR~0s are indistinguishable from FR~Is (e.g. BH mass, core and bolometric luminosities, host type and mass) and, in addition, FR~0s lie on the same correlation between JVLA core power and accretion power (estimated from [O~III] line and X-ray emission, \cite{baldi15,torresi18,baldi19}), defined by FR~Is: this clearly indicates a common central engine, where the jet is the main contributor to the whole bolometric power. At the extreme high energies, a few FR~0s (and candidates) have been detected in $\gamma$-ray band by the Fermi Telescope \cite{grandi16,stecker19,harvey20,paliya21} and have been proposed as possible sources of cosmic neutrinos and high energy cosmic rays \cite{jacobsen15,tavecchio18,merten21}. All these results foster the idea that the putative jets of FR~0s are possibly characterised by a relativistic bulk speed and that FR~0s may play a crucial role in galaxy evolution via a continuous low-regime jet feedback \cite{ubertosi21}, as supported by numerical simulations of low-power jets (e.g. \cite{massaglia16,mukherjee18,mukherjee20}).

Different scenarios have been suggested to account for the multi-band properties of FR~0s in relation with the other RG classes. FR~0s could be powered by a slowly spinning BHs \cite{baldi15,garofalo19}, limiting the amount of extracted energy available to launch and power the jet, thus resulting in compact, weak jets.  FR~0s could be early-phase FR~Is and will eventually evolve into the latter over a period of $\sim$10$^{8}$-10$^{9}$ years \cite{baldi18a}. FR~0 jets could be highly intermittent, switching on/off frequently during the course of their evolution \cite{baldi15,miraghae17,baldi18a}, resulting in the absence of prominent extended jets.

One of the best ways to probe the very inner parts of the ejection mechanism in this mostly-unexplored class of RGs is to study the pc-scale radio emission with very-long baselined imaging observations (VLBI technique), with the intention of finding the smoking-gun feature of their inability of launching fully-fledged  jet structure. At parsec scale, higher-resolution radio observations of a sample of FR~0s with world-wide VLBI, the American Very Long Baseline Array (VLBA) and European VLBI Network (EVN)  show resolved jets of a few pc for $\sim$80\% of the sample \cite{cheng18,cheng21}. The VLBI multi-epoch data and the symmetry of the radio structures indicates that the jet bulk speeds are mildly relativistic (between 0.08$c$ and 0.51$c$) with low bulk Lorentz factors (between 1.7 and 6) and large viewing angles. However, all previous VLBI-based studies focus on particularly bright FR~0s (flux densities higher than 50 mJy, a factor 10 higher than the typical FR~0 flux selection threshold \cite{baldi18a}). Therefore, a comprehensive study of pc-scale radio emission
in a significant and representative sample of FR~0s, moving to lower radio luminosities, is missing. Such a study would help to 
build an impartial view of the mechanisms of jet production in RGs, linking FR~Is to FR~0s. 

Together with VLBA and EVN, the {\it e}MERLIN  also provides a range of baseline lengths that permit unique studies of faint radio sources to be made over a wide range of spatial scales. In this work we will analyse the pc-scale radio emission of a sample of 15 low-power FR~0s with {\it e}MERLIN and EVN arrays.

This paper is organized as follows. In Sections~\ref{2} and \ref{3} we present the sample, observations and maps of the 5 FR~0s observed with the {\it e}MERLIN and 10 with EVN, respectively (Tab.~\ref{tab1}). The main properties of the pc-scale emission of FR~0s are presented in Sect.~\ref{results} and discussed in Sect.~\ref{discussion} in the context of disc-jet coupling in FR~0s in relation to other classes of RGs. We summarise the results and draw our conclusions in Sect.~\ref{s&c}.  A supplementary section \ref{app} provides the {\it e}MERLIN and EVN map parameters.

\begin{figure}	
\widefigure
\centerline{
\includegraphics[width=0.35\columnwidth]{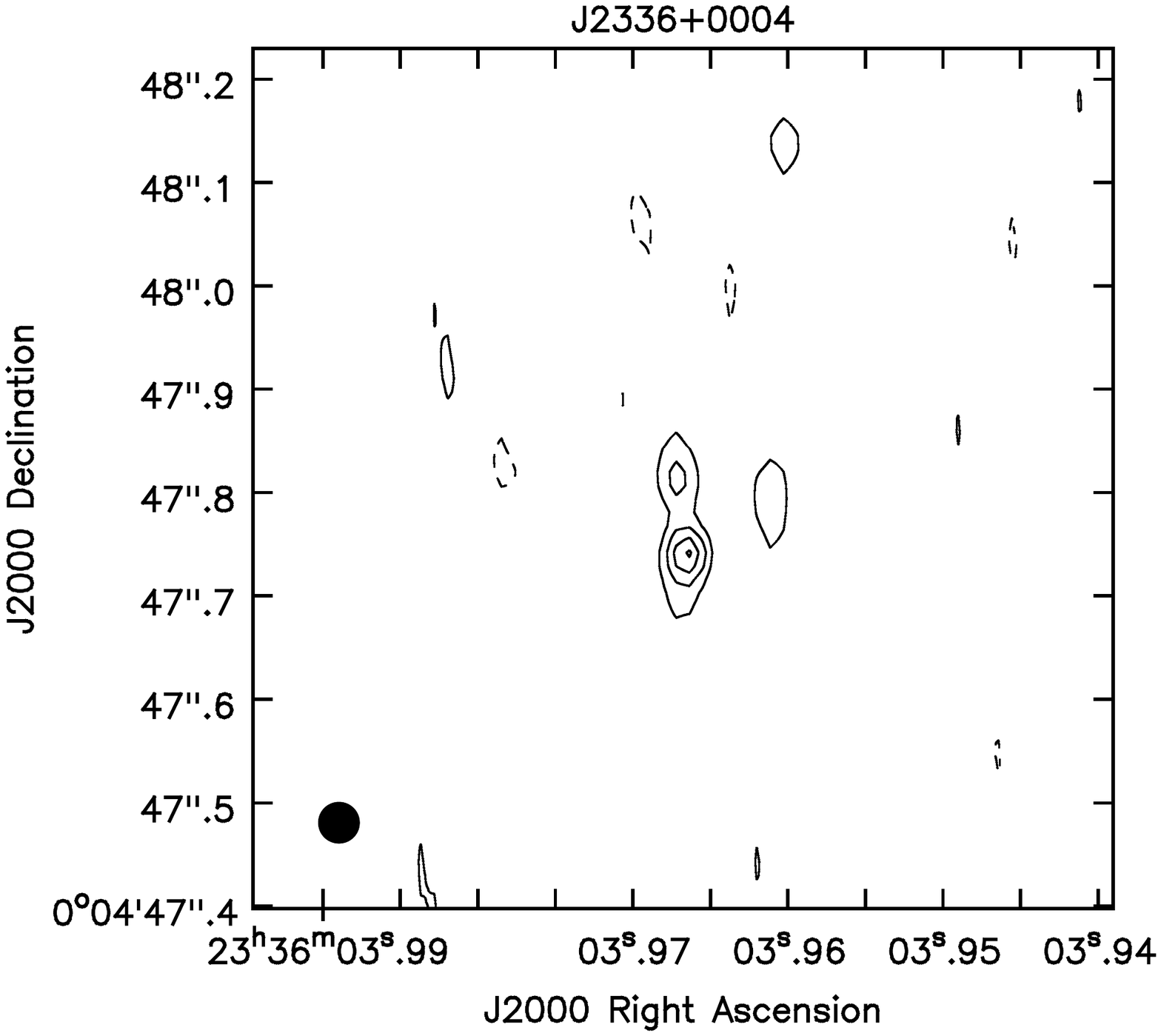}
\includegraphics[width=0.35\columnwidth]{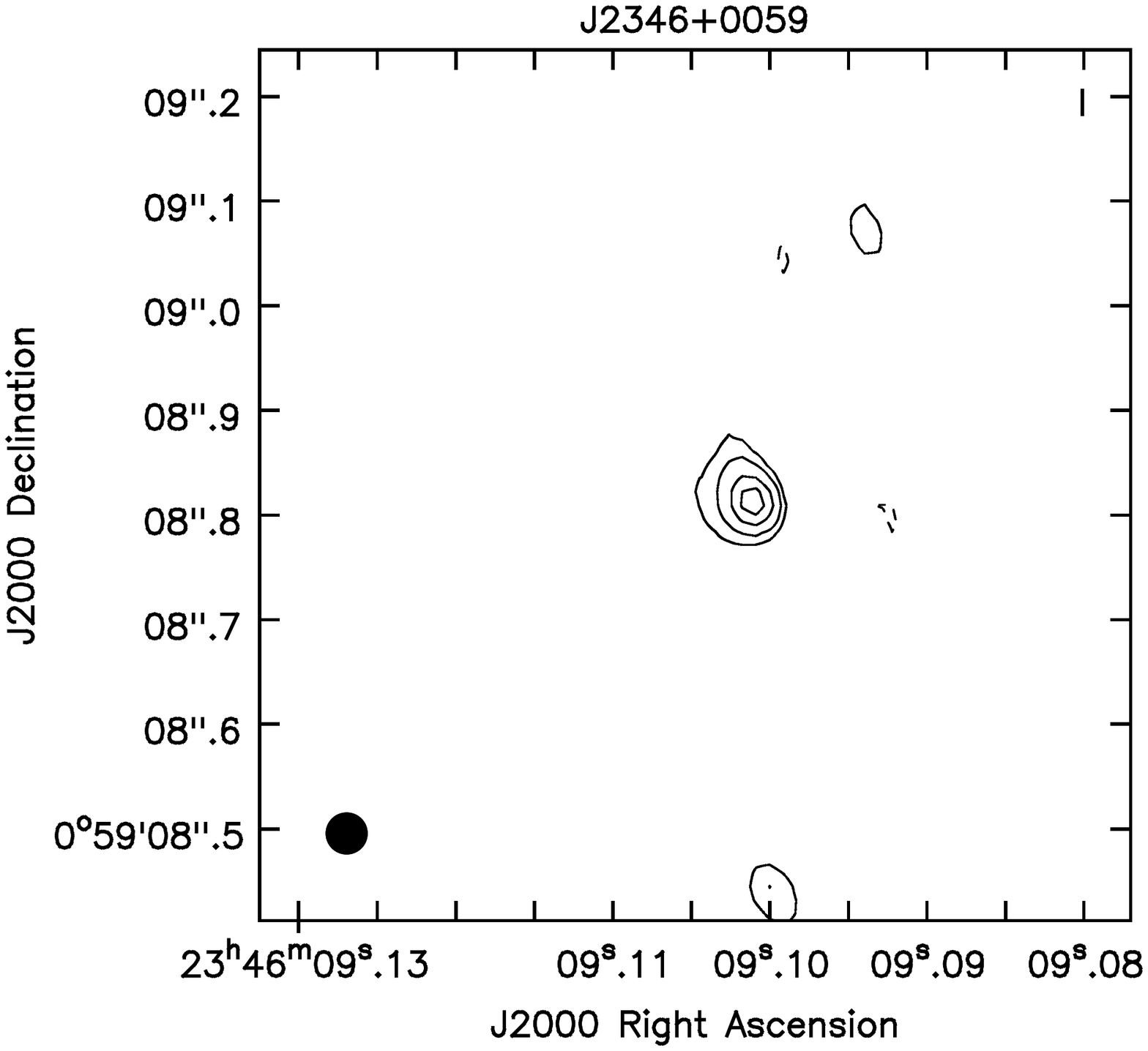}
\includegraphics[width=0.35\columnwidth]{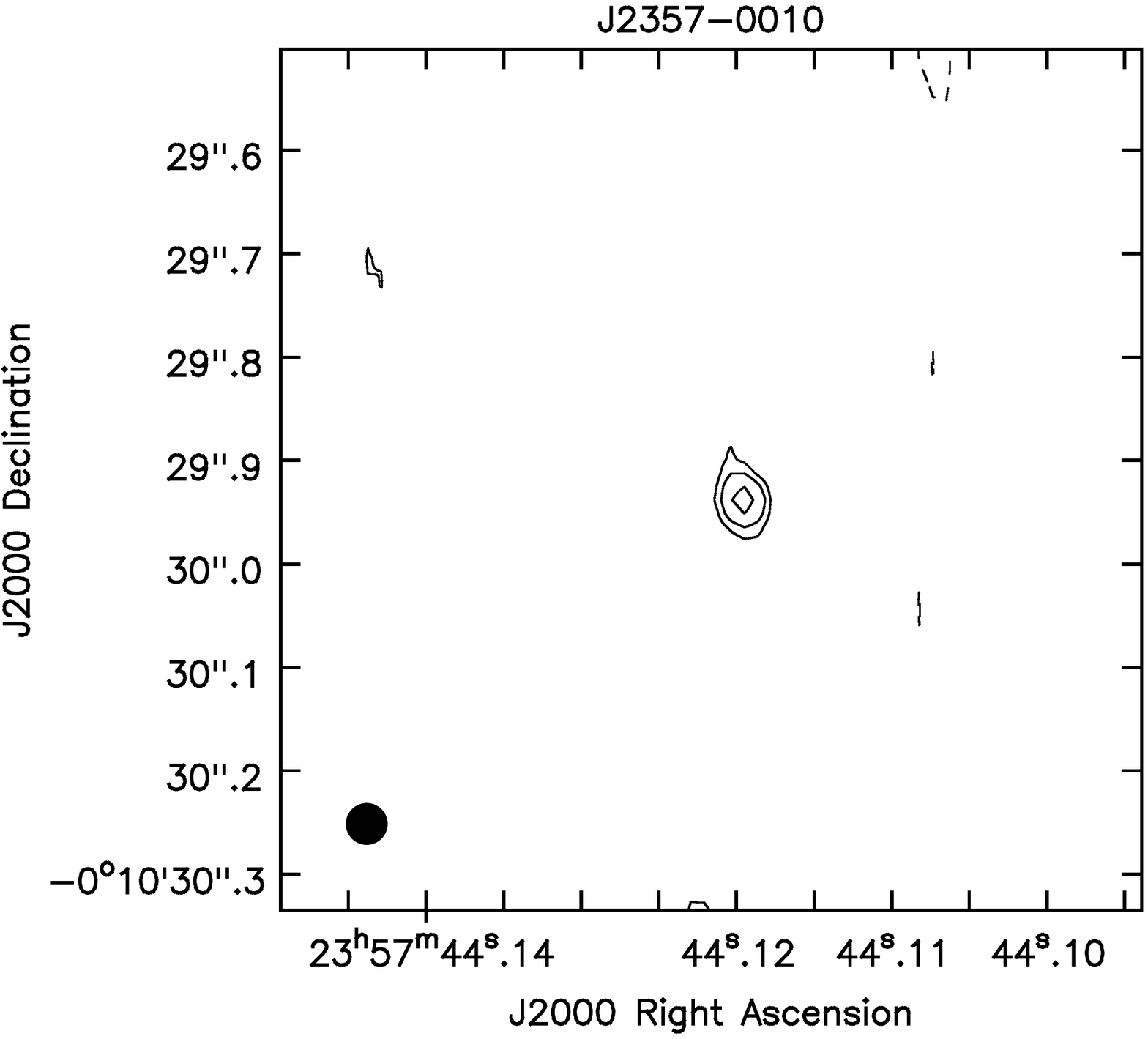}}
\vspace{-2.5cm}
\centerline{
\includegraphics[width=0.35\columnwidth]{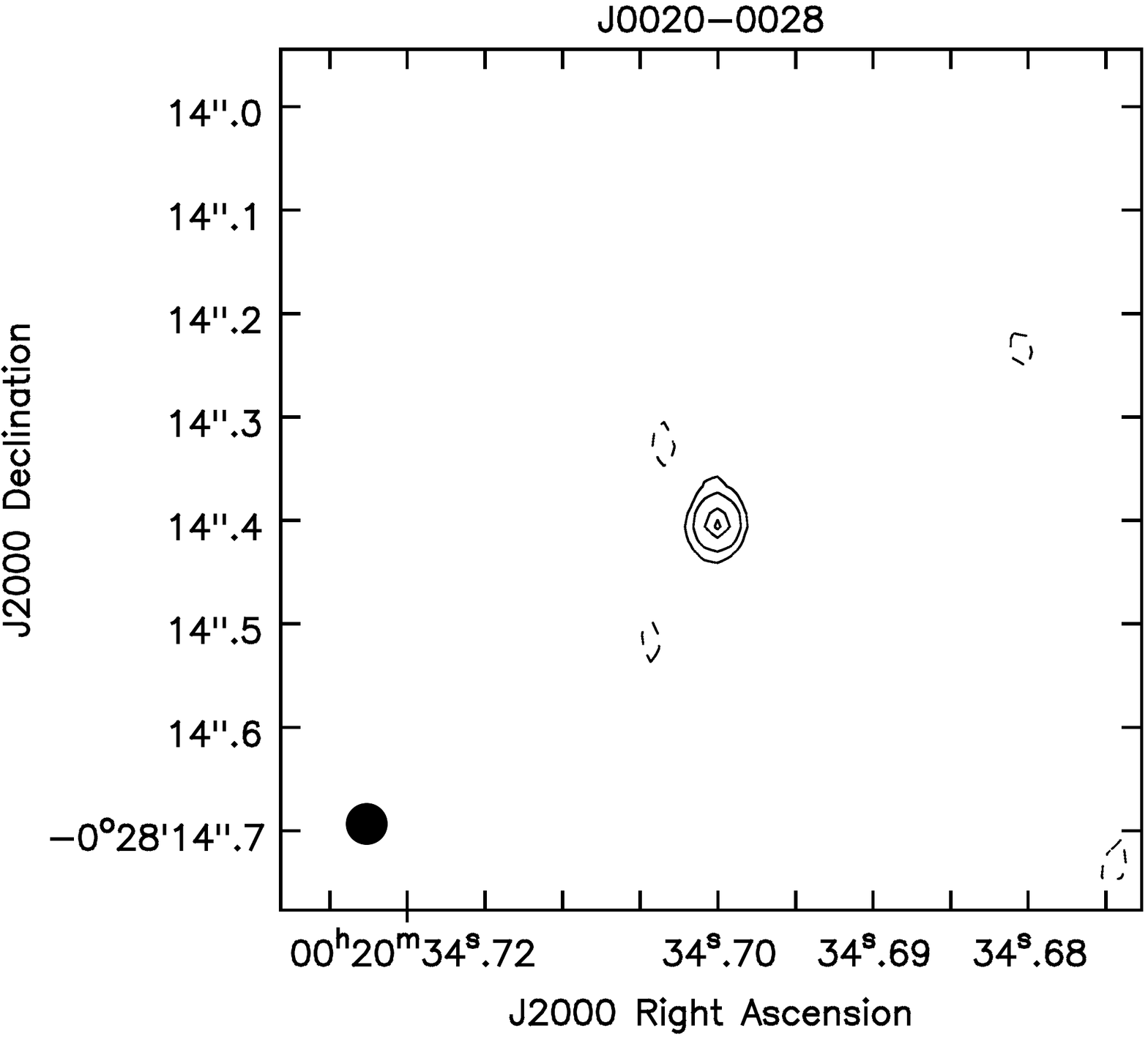}
\includegraphics[width=0.35\columnwidth]{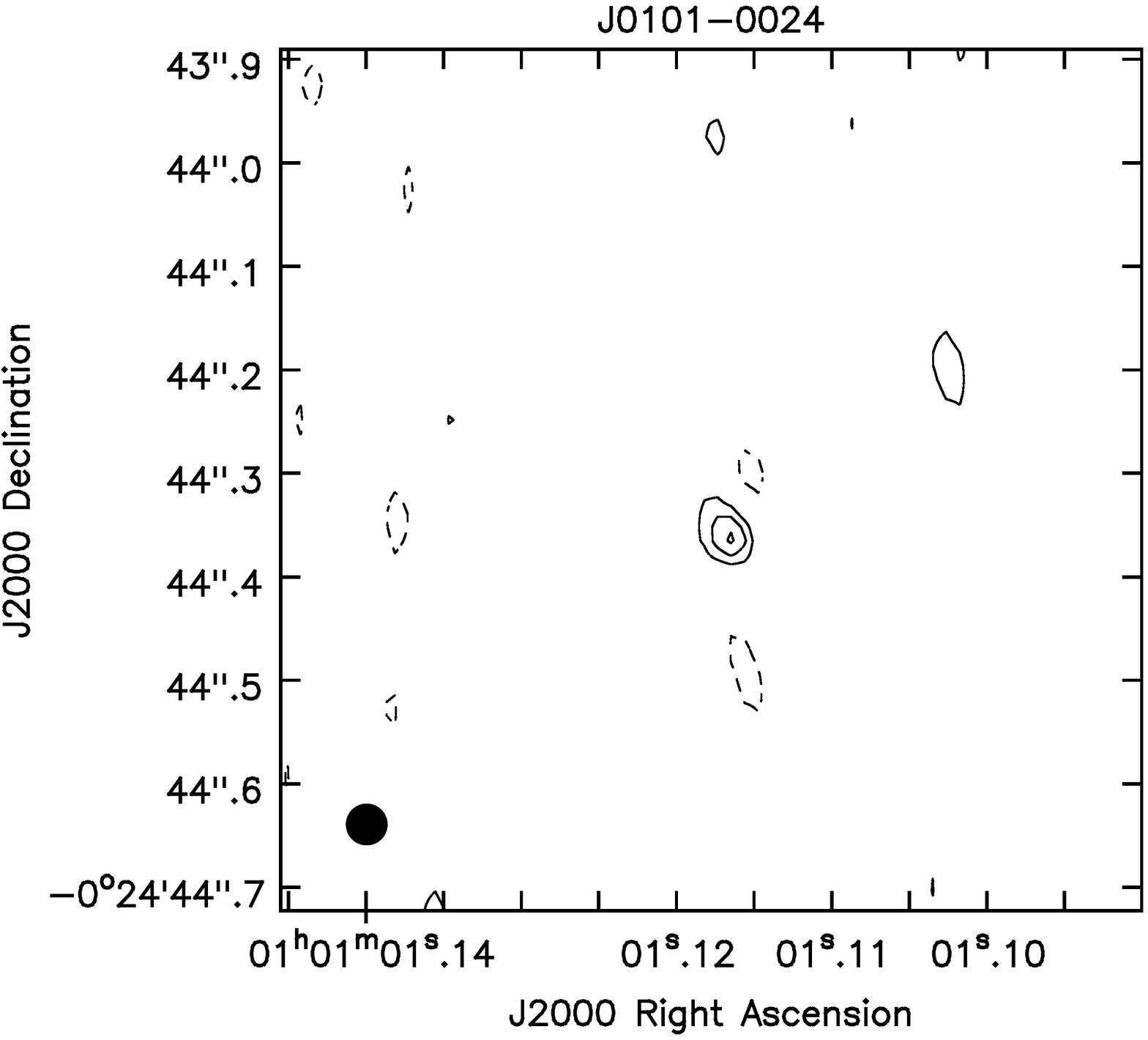}}
\vspace{-2.5cm}
\caption{The 5-GHz maps of the 5 FR~0s observed with the {\it e}MERLIN array. The filled area, shown at the bottom-left corner of the images, represents the restoring beam of the maps. Contours and beam parameters are tabled in Tab~\ref{contours}.\label{emerlin}}
\end{figure}

\end{paracol}
\begin{table}
\caption{Column Description: (1) name; (2) redshift; (3) phase calibrator of the {\it e}MERLIN and EVN observations; (4) BH mass (M$_{\odot}$); (5) [O~III] line luminosity (erg s$^{-1}$); (6)-(9) JVLA core luminosity (erg s$^{-1}$), morphology (SR slightly resolved or extended) and size (kpc) and radio spectral type (S steep, F flat, I inverted) with the 1.4-4.5 GHz JVLA spectral slope ($S_\nu \propto \nu^{\alpha}$).\label{tab1}}
\begin{tabular}{lcccccccc}
\toprule
\textbf{Name}	& \textbf{z} & \textbf{phase calib} & \textbf{$M_{\rm BH}$}  &   \textbf{L$_{\rm [O~III]}$} & \textbf{L$_{\rm JVLA}$} &  \textbf{morph} & \textbf{size}  & \textbf{spectra($\alpha_{\rm 1.4-4.5 \,\,GHz}$)} \\
\midrule
 \multicolumn{9}{c}{{\it e}MERLIN} \\
J2336+0004 & 0.076 & J2335-0131 & 8.7 &  40.28 & 39.31  & SR  &  1.3  & S(-1.0)\\
J2346+0059 & 0.093 & J2357-0152 & 8.3 & 39.96  &  $<$40.16  & SR & 0.5 & S(-0.56)\\
J2357-0010 & 0.076 & J2354-0019 & 8.8 & 40.26 &  39.29  & SR   & $<$0.3  & S(-0.67)\\
J0020-0028 & 0.072 & J0016-0015 & 7.6 & 39.97  & 38.93  &  disc & 2.8  & S(-0.89)\\
J0101-0024 & 0.097 & J0059+0006 &  8.4 & 40.39 & 39.64 & elongated & 1.5  & S(-0.45)\\
\midrule
 \multicolumn{9}{c}{EVN} \\
J0907+3257  &     0.049 & J0911+3349 &7.7 & 39.33 & $<$39.40 &  disc & 14 & S(-1.0)  \\ 
J0930+3413  &    0.042 & J0930+3503 & 8.4  & 39.93 & $<$39.59 & SR & & S(-0.49)\\
J0943+3614  &   0.022 & J0945+3534 & 7.9 & 39.85 & $<$40.3 & SR & & I(0.60)\\
J1025+1022  &    0.046 & J1025+1253 & 8.9 & 39.48 & 40.62 & SR & & F(0.12)\\
J1040+0910  &    0.019 & J1042+1203 & 8.3 & 39.54 & $<$39.19 & SR & & S(-0.60)\\
J1213+5044  &    0.031 & J1219+4829 & 8.7 & 40.12  & 40.09& two jets & 2 &  F(-0.33) \\
J1230+4700  &    0.039 & J1234+4753 & 8.4 & 40.00 & 40.13 & SR & & F(-0.29) \\
J1530+2705  &    0.033 & J1539+2744 & 8.2 & 39.71 & 39.93 & SR & & F(+0.19)\\
J1628+2529  &    0.040 & J1628+2247 & 8.5 & 39.65 & 39.87 & SR & & F(+0.11) \\
J1703+2410  &    0.031 & J1659+2629 & 8.8 & 39.66 & $<$38.85& two lobes & 9 & S(-0.63)\\
\bottomrule
\end{tabular}
\end{table}
\begin{paracol}{2}
\switchcolumn
 
\section{{\it e}MERLIN}\label{2}

\subsection{Sample, observations and maps}\label{sect2.1}

A sample of several thousands RGs selected by cross-correlating the optical SDSS and 1.4-GHz NVSS and
FIRST datasets (hereafter the SDSS/NVSS sample) has been recently created \cite{best05b}, which mostly ($\sim$80\% of the sample) consists of compact unresolved objects at 5$\arcsec$-scale with FIRST \cite{baldi10}. This sample includes compact RGs with FIRST flux densities $>$ 5 mJy up to z $\sim$0.3, covering the range of radio luminosity $\sim$ 10$^{22}$--10$^{26}$ W Hz$^{-1}$. Based on a JVLA follow-up of 12 sources in L and C band, \cite{baldi15} classified seven FR~0s based on their optical and radio properties at 1.4, 4.5 and 7.5 GHz. We proposed {\it e}MERLIN observation of this sample, but only five FR~0s (named {\it e}MERLIN FR~0 sample) have been observed in C band (CY4213 project), see Table~\ref{tab1} (upper part). 

The five sources are randomly selected from the seven FR~0s studied with the JVLA \cite{baldi15}. Two sources appear extended on a few kpc at sub-arcsec resolution with JVLA: J0020-0028 show a twin 1$\arcsec$ structure, probably associated with the dusty disk of an edge-on lenticular galaxy and J0101-0024 displays an 0.8$\arcsec$ elongated bent structure. The five FR~0s have  FIRST flux densities in the range 5-25 mJy and are characterised by a steep radio spectra at 1.4-4.5 GHz, but a spectral flattening emerges at  higher frequencies (towards 7.5 GHz). They are all hosted in red massive ETGs with BH masses $>$10$^{7.5}$ M$_{\odot}$.

The {\it e}MERLIN observations of the five FR~0s were carried in December 2016 with 6 telescopes (without the Lovell telescope) spread across the UK (having a maximum  baseline length of 217 km), reaching a nominal resolution of $\sim$40 mas at the C band.  The observations were centered at 5 GHz with a bandwidth of 512 MHz (4 spectral windows) and correlated at Jodrell Bank Observatory. The observations were divided in two observing blocks, where two or three targets are stored together with their phase and flux calibrators.  We scaled the flux density  using $\sim$15 min scans of 3C~286 and $\sim$10 min scan of the bandpass calibrator OQ~208 at the beginning of each run. Then we tracked the complex antenna gains using regular $\sim$2.5 min scans of the bright phase reference source (see list in Tab~\ref{tab1}) which we interleaved between 6.5 min scans on the target field, with a total target time-on-source of $\sim$2 hours.

We calibrated the dataset with the {\it e}MERLIN CASA data reduction pipeline \cite{moldon21}, which is intended to automate the procedures required in processing and calibrating radio astronomy data from the {\it e}MERLIN correlator. After removing the low-level RFIs with the AOFlagger software \cite{offringa12}, we  also performed a standard calibration procedure by  fitting the offsets in delays using the CASA task ‘fringefit’, by calculating the bandpass and phase solutions\footnote{See \cite{baldi21a} for more details on the {\it e}MERLIN pipeline and the comparison with standard AIPS data reduction and imaging.}. The complex gains (phase and amplitude)  were improved with several runs of self-calibration on the gain calibrators and then applied to all the targets. After a further inspection of the target data to check the quality and remove possible bad scans, the calibrated datasets were imaged (Stokes I) within the CASA environment using the task 'tclean' with the deconvolver mode 'mtmfs', which performs a   multiscale, multi-frequency synthesis algorithm \cite{rau11}, with a cell size of 0.013$\arcsec$ and natural weighting. For the targets with flux densities higher than 5 mJy, we carried out a few rounds of self-calibration in phase and a final one in phase and amplitude, using 1-2 min integration times and using a 3$\sigma$ minimum threshold for valid solution. 

To analyse the source parameters in the radio maps, we used ‘imfit’, part of the CASA ‘viewer’, which fits
two-dimensional Gaussians to an intensity distribution on a region selected interactively on the map, providing the
position, the deconvolved size, peak flux density, integrated flux density, and position angle (PA) of the compact source (all listed in Tab.~\ref{contours}). Whilst this procedure is valid for compact components, to estimate the total brightness of an extended component is by interactively marking the region around the irregular shape of the
source with CASA 'viewer'. Figure~\ref{emerlin} displays the the 5-GHz maps of the five FR~0s with angular resolution (restoring beam) of $\sim$40 mas and rms between 40 - 80 $\mu$Jy beam$^{-1}$.

The five sources are detected with flux densities ranging from 0.4 and 2.7 mJy beam$^{-1}$ at 5 GHz (see Tab~\ref{emerlin_flux}). The sources appear all slightly resolved, except J2336+0004 which shows a weak second component with PA $\sim$ 7$^{\circ}$. The angular lengths, estimated from the convolved-beam major axis, considered as the largest angular scale of the source, are of $\sim$50 mas, apart from J2336+0004 which is 0.15$\arcsec$: the linear physical scales correspond to $\sim$70-240 pc.  All the targets, even the two JVLA resolved sources (J0020-0028 and J0101-0024),  do not show significant evidences of jetted extended morphologies at parsec scale.

\end{paracol}
\begin{paracol}{2}
\switchcolumn

\end{paracol}
\begin{table}
\caption{Column Description: (1) name; (2)-(5) core peak flux density (mJy beam$^{-1}$),  core brightness temperature (units of 10$^{4}$ K), total core flux density (mJy) and its core luminosity (erg s$^{-1}$) from {\it e}MERLIN maps at 5 GHz; (6)-(7) radio morphology (TS two-sided, OS one-sided, SR slightly resolved, UR unresolved) and size (pc); (8)  core peak flux density (mJy beam$^{-1}$) from the JVLA maps \cite{baldi15}; (9)-(10) core peak flux density (mJy beam$^{-1}$) and total emission from the combined JVLA-{\it e}MERLIN maps; (11)-(13) PA (degree), radio morphology and size (kpc); (14) jet sidedness.\label{emerlin_flux}}
\begin{tabular}{lcccccc|c|cccccc}
\toprule
\textbf{Name}	& \textbf{F$_{\rm core}$} & \textbf{T$_{\rm B}$}  & \textbf{F$_{\rm tot}$} & \textbf{L$_{\rm core}$ }& M	& size & \textbf{F$_{\rm JVLA}$}  &   \textbf{F$_{\rm comb \,core}$} &  \textbf{F$_{\rm comb \,tot}$}  & \textbf{PA} &  \textbf{M}	& \textbf{size}  &   \textbf{jet ratio}\\
\midrule
J2336+0004 &  0.64$\pm$0.07   & 2.0 &  1.1   &  40.28 & OS? & 240  & 1.96 & 1.03$\pm$0.04 & 8.5 & 46 & TS & 1.1  & 1.8$\pm$0.2 \\ 
J2346+0059 &  2.66$\pm$0.14  & 8.1 &  3.66 & 39.96   & SR            & 100 & 9.16 & 7.27$\pm$0.14 & 13.3 & 29 &TS &  0.88  & 1.1$\pm$0.1 \\ 
J2357-0010 &  0.46$\pm$0.16  & 1.4 &  0.47 &  40.26  & UR            &   70  & 1.86 & 0.91$\pm$0.03 & 2.0 & 36  &TS &  0.46  & 1.1$\pm$0.1 \\ 
J0020-0028 &  0.64$\pm$0.05  & 2.0 &  0.72 & 39.97   & UR            &    74      & 2.62 & 0.74$\pm$0.04 & 1.2 & -45 & SR & 0.30  & \\  
J0101-0024 &  0.38$\pm$0.06  & 1.2 &  0.57 & 40.39   & SR            &     120   & 2.5 & 0.82$\pm$0.03 & 3.0 &  29/38 & TS &  1.5  & 1.6$\pm$0.2 \\ 

\bottomrule
\end{tabular}
\end{table}
\begin{paracol}{2}
\switchcolumn

\subsection{Combining JVLA and {\it e}MERLIN data}

Since we own the calibrated C-band JVLA observations of these five sources  (carried out in 2012-2013),  we combined the calibrated {\it e}MERLIN visibilities with those from  JVLA in the same band to perform a hybrid map, obtained with long and short baselines. To match the overlapping bandwidth of two datasets, we exported the first two spectral windows of the {\it e}MERLIN visibilities (centered at  4816.5 and 4944.5 MHz) and the last two spectral windows of the JVLA visibilities (centered at 4819 and  4947 MHz). First, we performed a first shallow re-weight of each individual visibility by its rms value  with the CASA task 'statwt' to contribute equally in the combined dataset  given that two radio arrays both have similar image noises and antennas sizes. This process is equivalent to combining the data sets prior to Fourier transformation into the sky-plane. Both the CASA measurement sets files were imaged simultaneously with task 'tclean', using  multiscale parameters [0,5,10] to give the best compromise between diffuse structure and unresolved sources. A further issue to solve is that the combined JVLA-{\it e}MERLIN beam is not a regular Gaussian due to the presence of wide shoulders owing to the short spacings provided by the JVLA. Therefore, following the procedure discussed by \cite{radcliffe21}, we reduced the possible flux density offsets between the combined maps and the original ones by deconvolving the combined dirty maps to deeper thresholds than those reached in the individual JVLA and {\it e}MERLIN maps (see Table for the typically lower rms obtained), and  re-weighing the data so that the resultant beam more closely represents a Gaussian. Systematic effects and complications, caused by an extreme deconvolution, can be mitigated with continuous prudent iterations \cite{radcliffe22}.

\end{paracol}
\begin{figure}[ht]
\widefigure
\centerline{
\includegraphics[width=0.35\columnwidth]{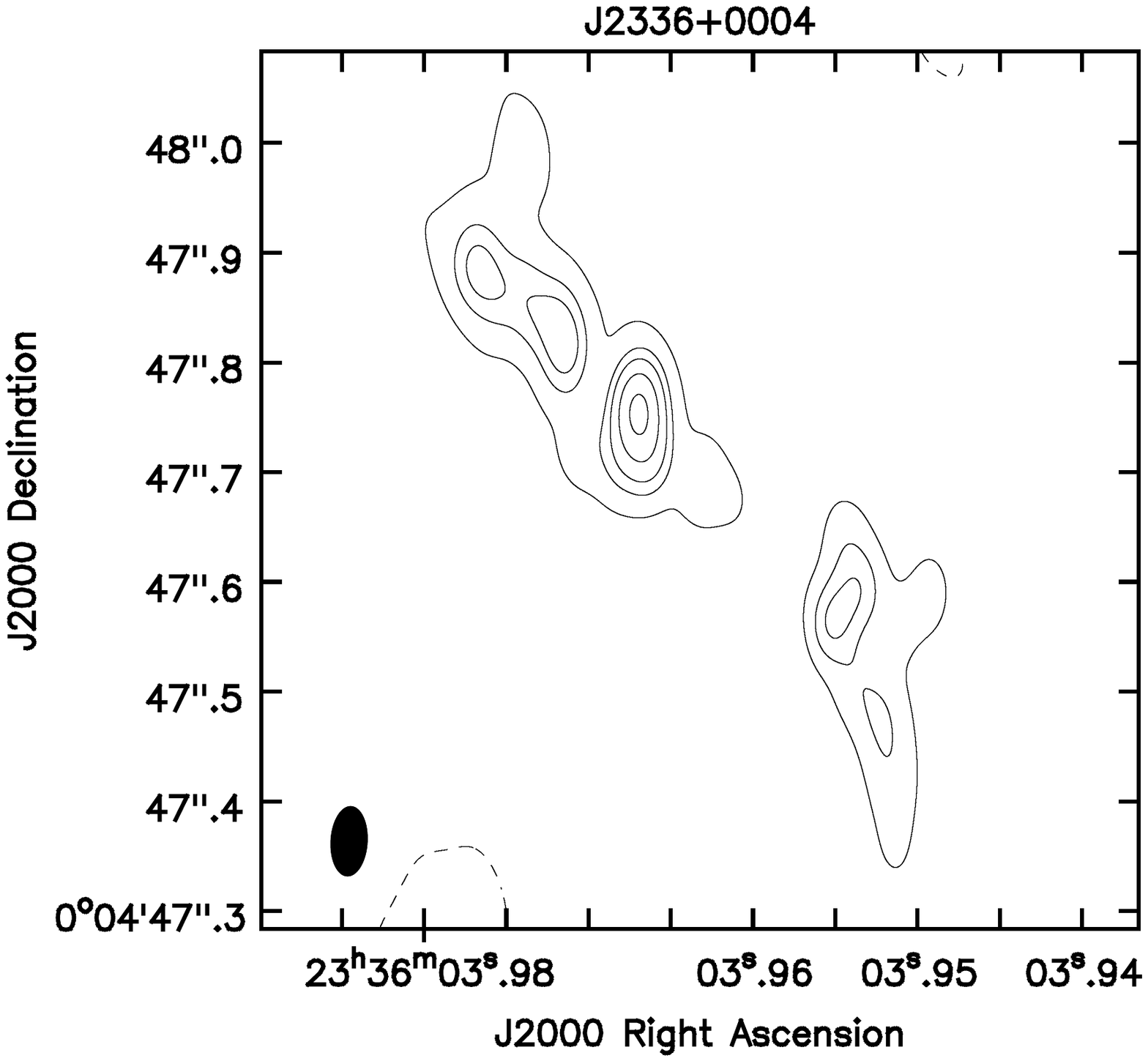}
\includegraphics[width=0.35\columnwidth]{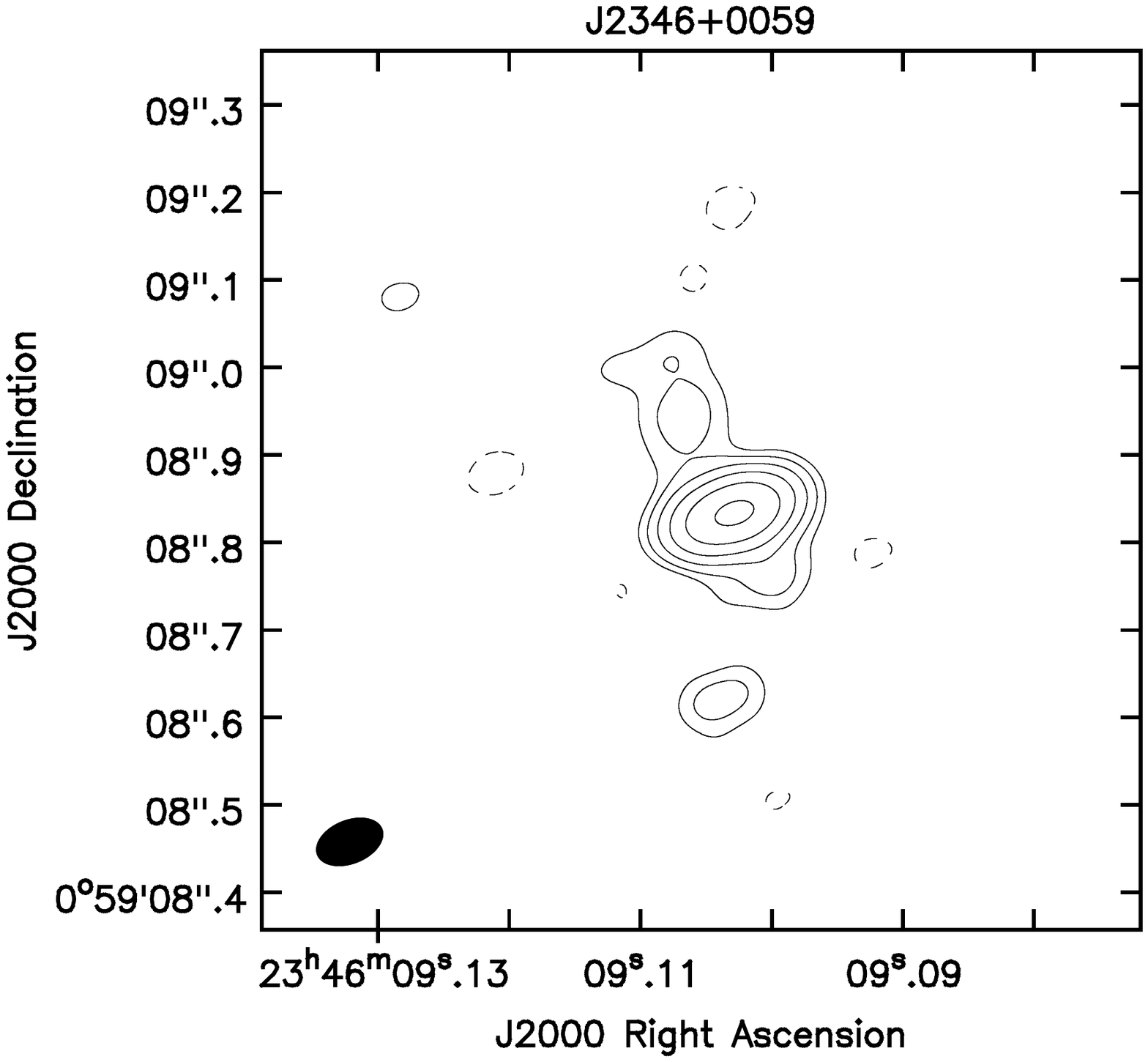}
\includegraphics[width=0.35\columnwidth]{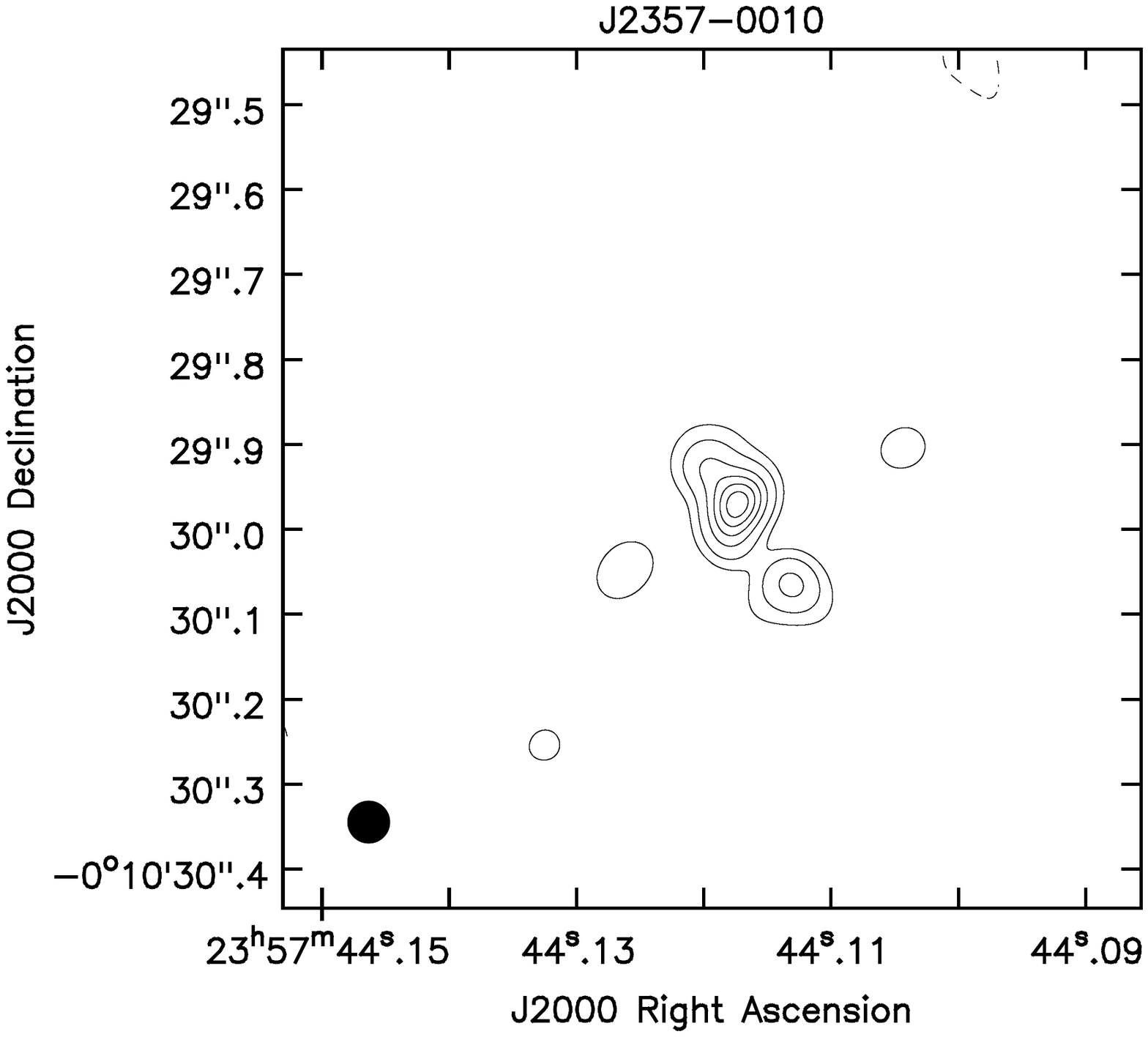}}
\vspace{-3cm}
\centerline{
\includegraphics[width=0.35\columnwidth]{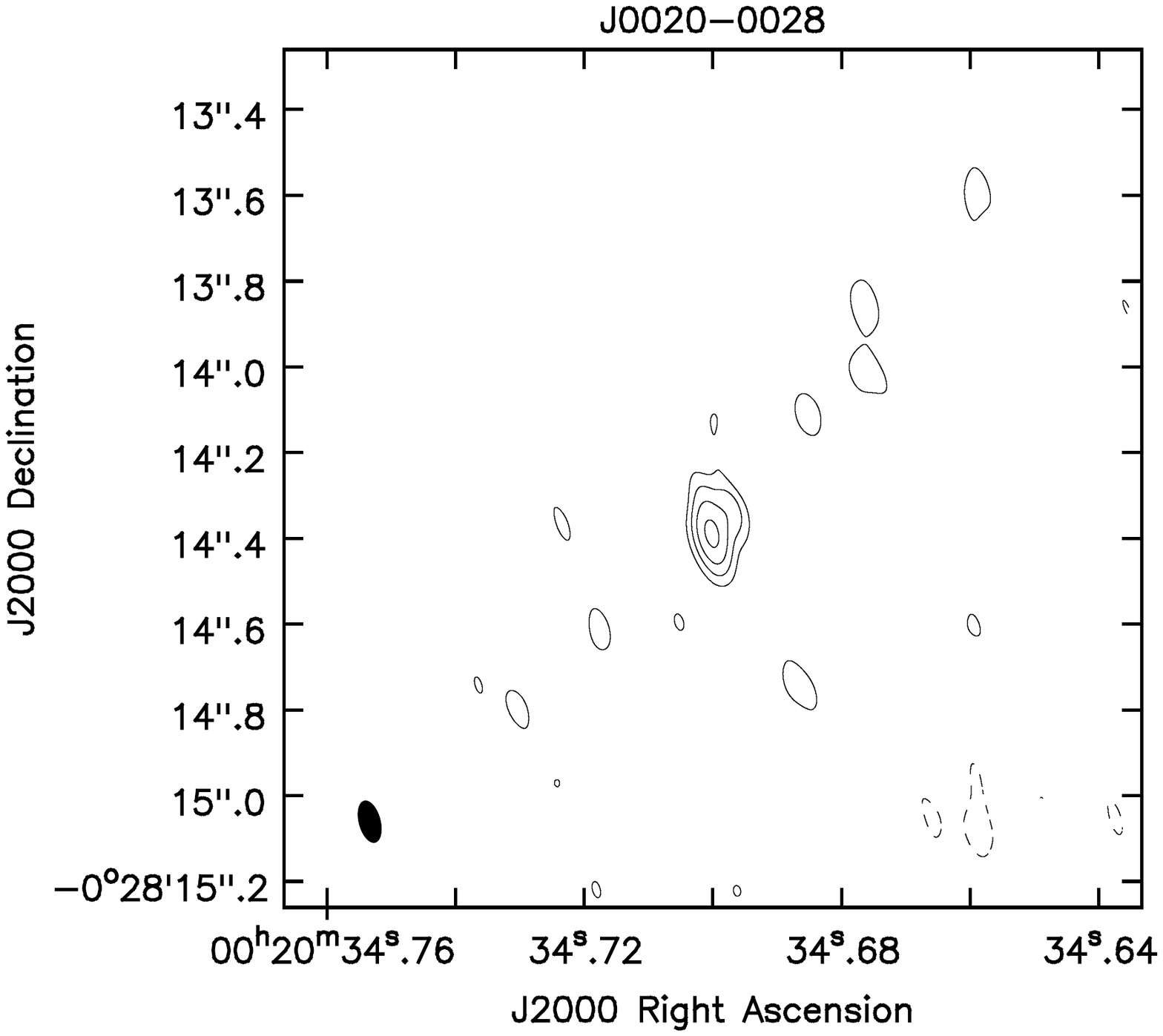}
\includegraphics[width=0.35\columnwidth]{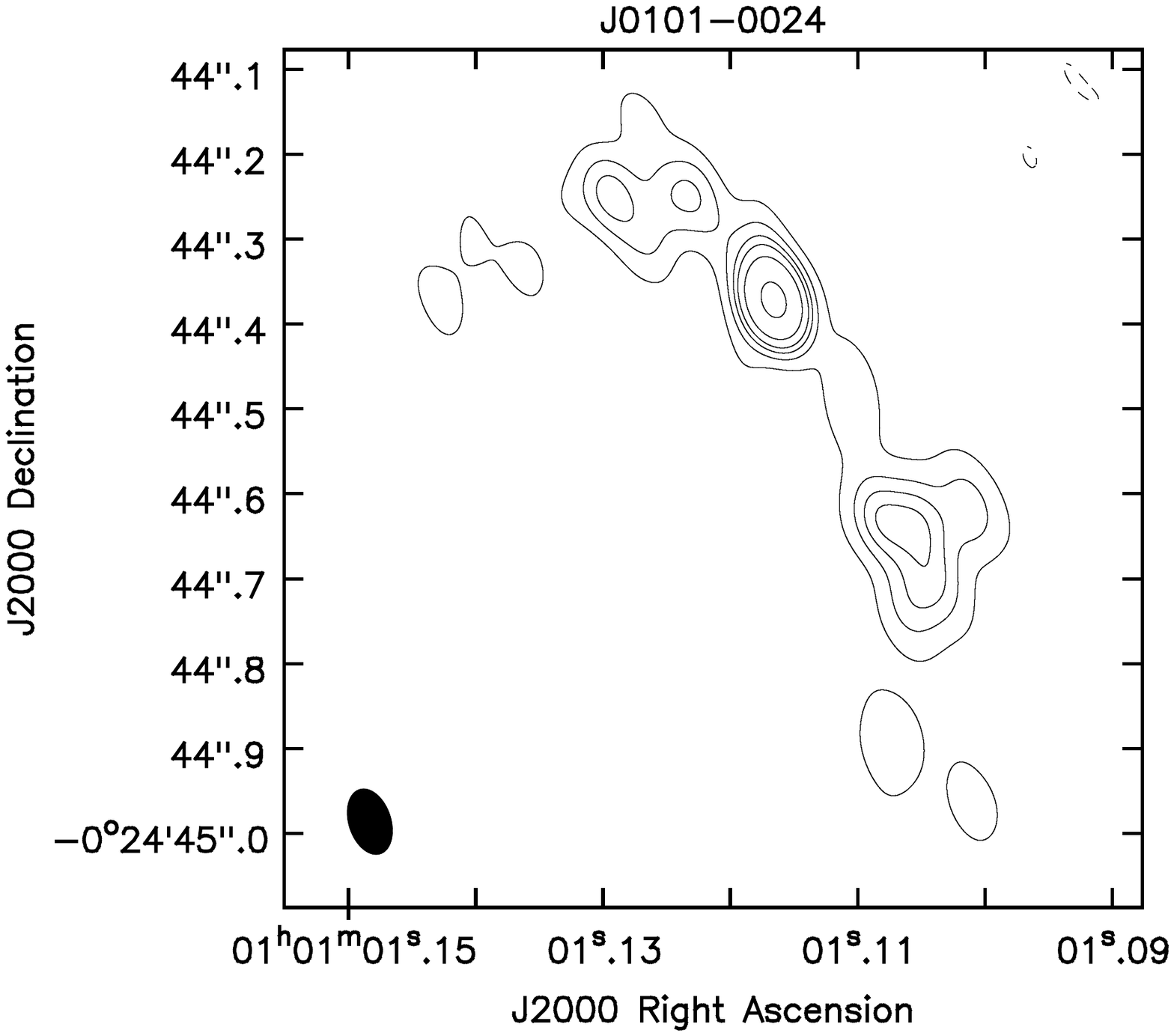}}
\vspace{-3cm}
\caption{The 4.9-GHz maps of five FR~0s, obtained by combining {\it e}MERLIN and JVLA visibilities. The filled area, shown at the bottom-left corner of the images, represents the restoring beam of the maps. Contours and beam parameters are tabled in Tab~\ref{contours}.\label{combined}}
\end{figure}  
\begin{paracol}{2}
\switchcolumn

Figure~\ref{combined} presents the maps of the combined JVLA-{\it e}MERLIN datasets for the 5 FR~0s centered at 4.9 GHz with an angular resolutions less than 0.1$\arcsec$ (see Tab~\ref{contours}). Their rms (a few tens of $\mu$Jy) are  lower than that of the {\it e}MERLIN maps because deep self-calibration allowed to reduce the noise levels and to reduce the problem of the irregular combined JVLA-{\it e}MERLIN beam\footnote{As exception, the combined map of J2336+0004 has a higher rms than that of the single {\it e}MERLIN map because the JVLA phase calibration of this target was poor \cite{baldi15}.}. All the targets appear extended, apart from J0020-0028 where only a core with scattered components along the radio axis (PA $\sim$ 45$^{\circ}$) has been detected\footnote{The extended radio emission, consistent with the 1.4-GHz JVLA maps (PA$\sim$40$^{\circ}$, \cite{baldi15}) is visible in the  dirty image but after the deconvolution only the core and scattered point sources remain in the final map.}. Two-sided jetted morphologies with extents ranging  (0.2-0.8$\arcsec$) $\sim$0.3-1.5 kpc have been resolved, with core components typically of 1 mJy beam$^{-1}$, apart from J2346+0059 which outstands for its brightness, 7.3 mJy beam$^{-1}$. The twin jets detected in J2336+0004 have a  PA of 46$^{\circ}$, significantly different from what has been measured from the marginally one-sided pc-scale extent (PA $\sim$7$^{\circ}$) detected with the {\it e}MERLIN array singularly. J0101-0024 displays two-sided structures with lobes, with a slightly convex bending of the jets (PA moving from 38$^{\circ}$ to 29$^{\circ}$  from the north to the south jets), consistent with the elongated gentle curvature noted in the JVLA map \cite{baldi15}.

{\it e}MERLIN resolves the core component, probably related to the jet basis, for the five FR~0s with a core dominance F$_{\rm core}/F_{\rm NVSS}$ of typically $\sim$6 \%. The apparent absence of pc-scale jets in the {\it e}MERLIN maps is probably due to the intrinsic weakness of their radio brightness. The kpc-scale jetted structures detected in the combined maps consist of 40-80\% of the total emission and the pc-scale core components detected in full-resolution with the {\it e}MERLIN  consist of 10-50 \% of the total emission detected in the combined maps.

\section{EVN}
\label{3}
\subsection{Sample, observations and maps}
\end{paracol}
\begin{table}
\caption{Column Description: (1) name; (2)-(6) core peak flux density (mJy beam$^{-1}$), core brightness temperature (K), total core flux density (mJy) and total emission from the entire structure (mJy), and its core luminosity (erg s$^{-1}$) from EVN maps at 1.66 GHz; (7)-(9) PA (degree), radio morphology (TS two-sided, OS one-sided, SR slightly resolved, UR unresolved) and size (pc); (10) jet sidedness. For J1040+0910 the core location within the structure is ambiguous and we mark with ${*}$ its core flux densities and luminosity, estimated assuming its most south brightest component as core. \label{evn_flux}}
\begin{tabular}{lccccccccc}
\toprule
\textbf{Name}	& \textbf{F$_{\rm core}$} &   \textbf{T$_{\rm B}$} & \textbf{$F_{\rm int}$} &  \textbf{F$_{\rm tot}$} &\textbf{L$_{\rm core}$} & \textbf{PA} & \textbf{morph}	&  \textbf{size}  &   \textbf{jet ratio}\\
\midrule
J0907+3257  & 7.61$\pm$0.14  & 1.4$\times$10$^{7}$ & 13.48$\pm$0.38 & 19.5 & 39.33 & 60 & TS & 133  & 6.5$\pm$1.0 \\ 
J0930+3413  &  4.46$\pm$0.08 & 1.4$\times$10$^{7}$ & 5.66$\pm$0.18  & 7.8 & 39.93 & -86 & OS &  26 & $>$2 \\ 
J0943+3614  &     97.8$\pm$0.9 & 3.9$\times$10$^{8}$ & 128.3$\pm$2.0 & 144 & 39.85 & & SR  &   5.4  & \\ 
J1025+1022  &     84.8$\pm$0.5 & 1.6$\times$10$^{8}$ & 105.7$\pm$1.0 & 106  & 39.48 & & UR & 50  &  \\ 
J1040+0910  &    2.29$\pm$0.02$^{*}$ & 5.6$\times$10$^{6}$$^{*}$  & 4.7$\pm$0.3$^{*}$  & 9.5 & 39.54$^{*}$ & -12/24 & ? &  29 pc &  1.4-3  \\ 
J1213+5044  &   50.1$\pm$0.7 & 3.6$\times$10$^{7}$ & 53.8$\pm$1.3 & 55 &  40.12 & -25 & OS & 51 pc & 2.8$\pm$0.2 \\ 
J1230+4700  &    36.77$\pm$0.35 & 4.5$\times$10$^{7}$ & 37.22$\pm$0.64 & 65.8 & 40.00 & -7 & TS &  128  & 1.5$\pm$0.1 \\ 
J1530+2705  &    52.8$\pm$1.4 & 9.3$\times$10$^{7}$ & 57.4$\pm$2.8 & 58.8 & 39.71 & 62  & OS &  22 & $>$1 \\ 
J1628+2529  &    21.33$\pm$0.34 & 3.9$\times$10$^{7}$  & 26.67$\pm$0.71 & 27.0 & 39.65 & & UR &  20  &  \\  
J1703+2410  &    4.15$\pm$0.06 & 1.2$\times$10$^{7}$ & 4.72$\pm$0.07  &  7.5 & 39.66 & 70 & TS &  26  & 1.0$\pm$0.1 \\ 
\bottomrule
\end{tabular}
\end{table}
\begin{paracol}{2}
\switchcolumn


\end{paracol}
\begin{figure}[ht]
\widefigure
\centerline{
\includegraphics[width=0.3\columnwidth]{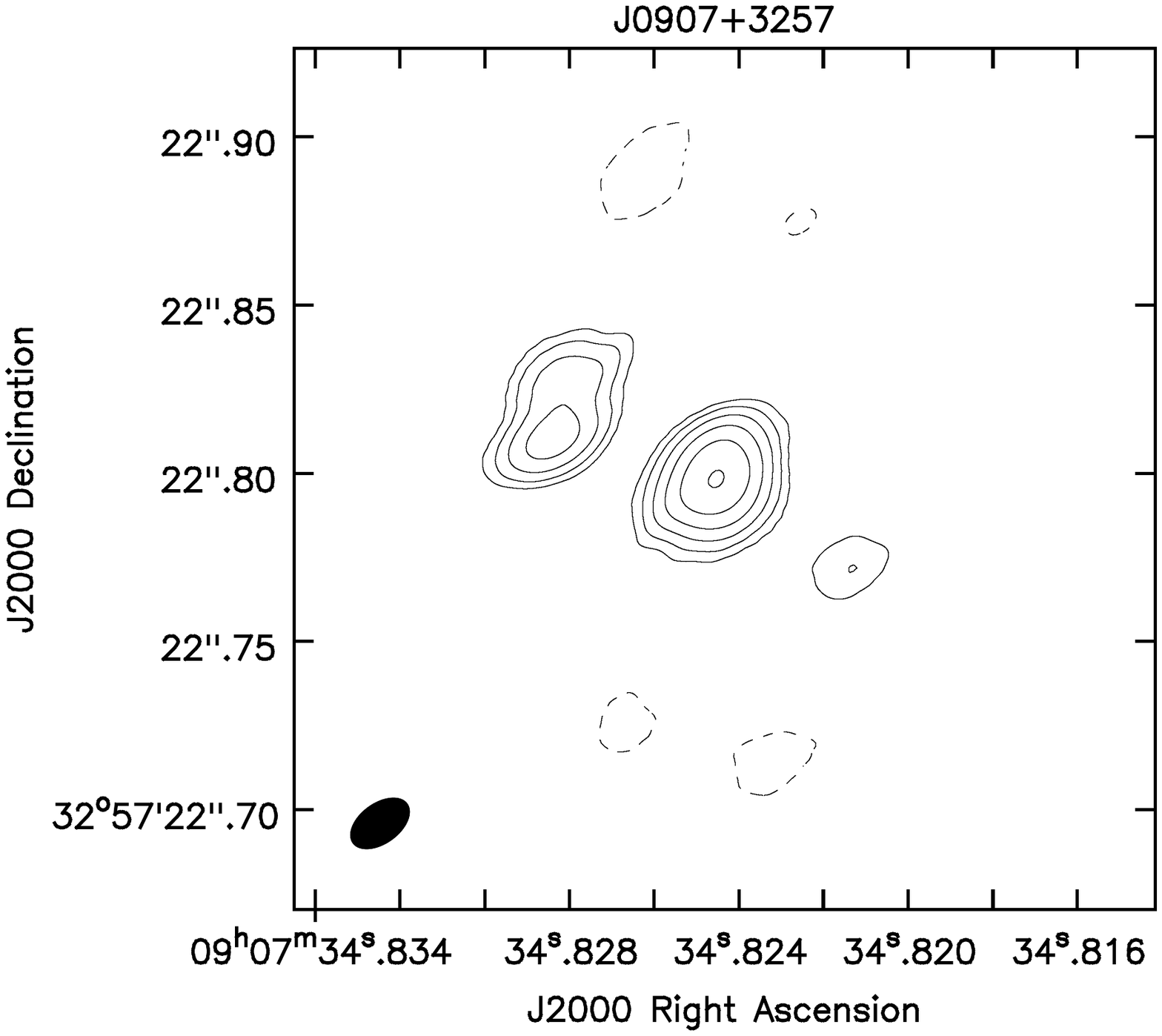}
\hspace{-1cm}
\includegraphics[width=0.3\columnwidth]{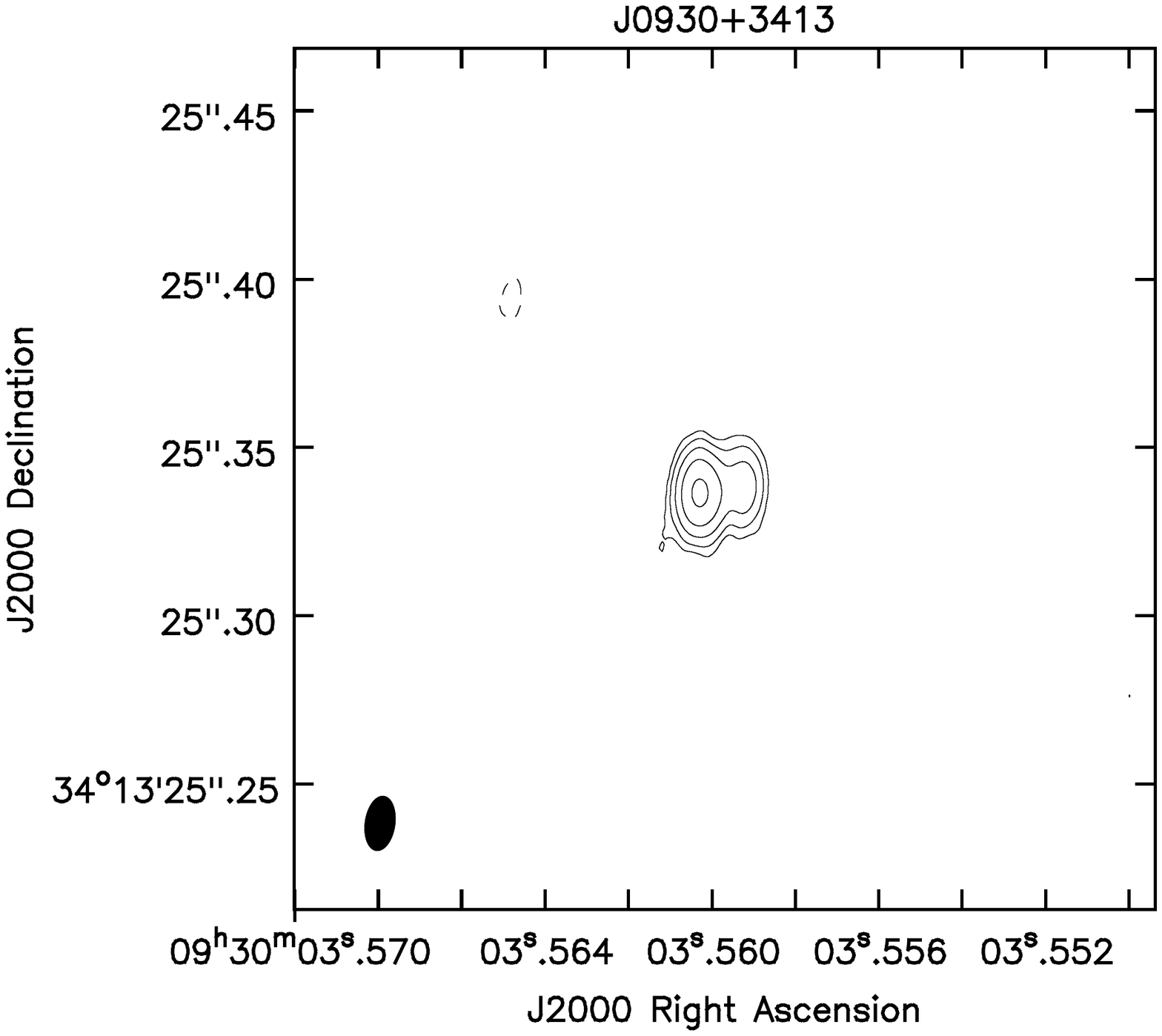}
\hspace{-1cm}
\includegraphics[width=0.3\columnwidth]{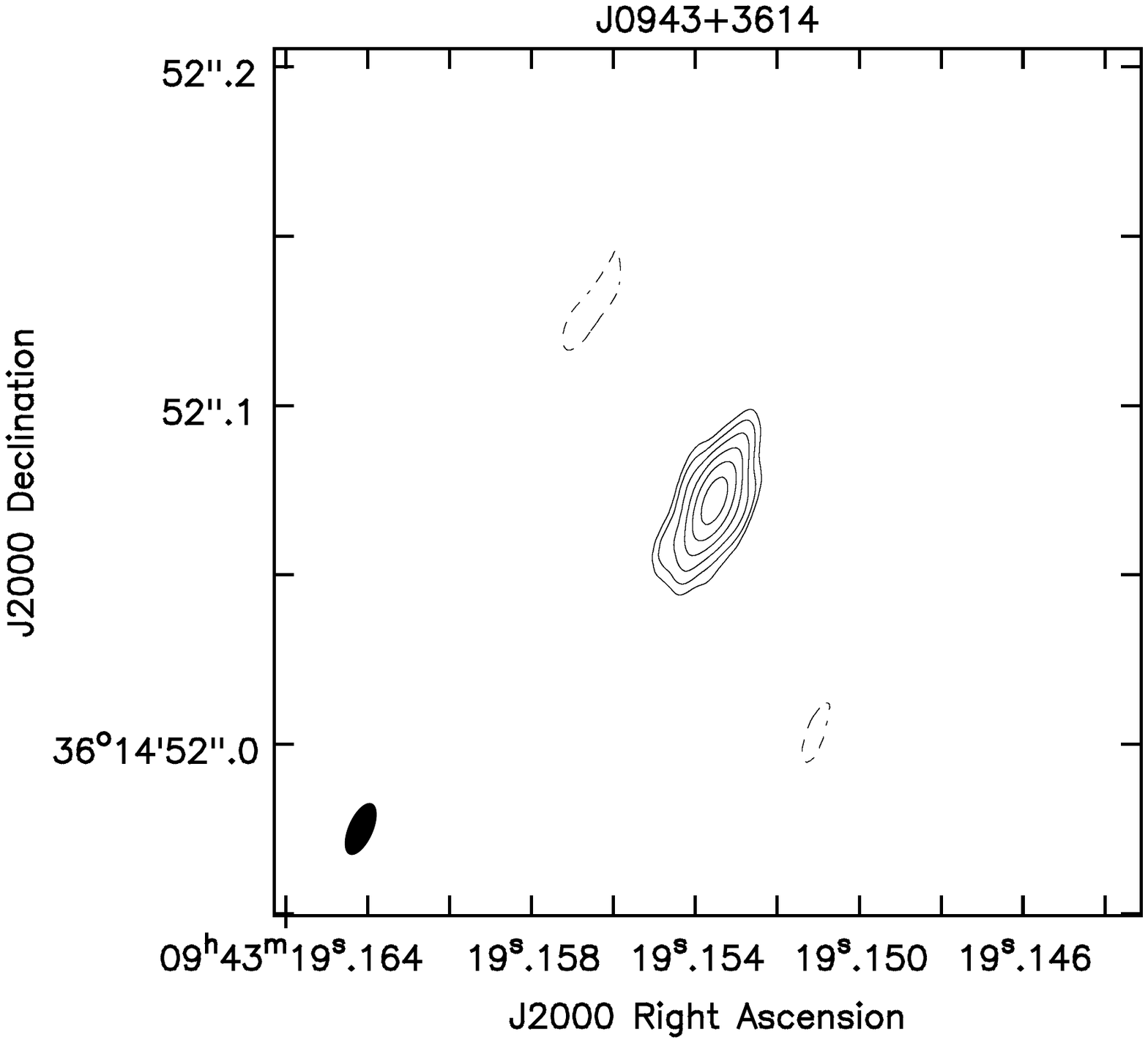}
\hspace{-1cm}
\includegraphics[width=0.3\columnwidth]{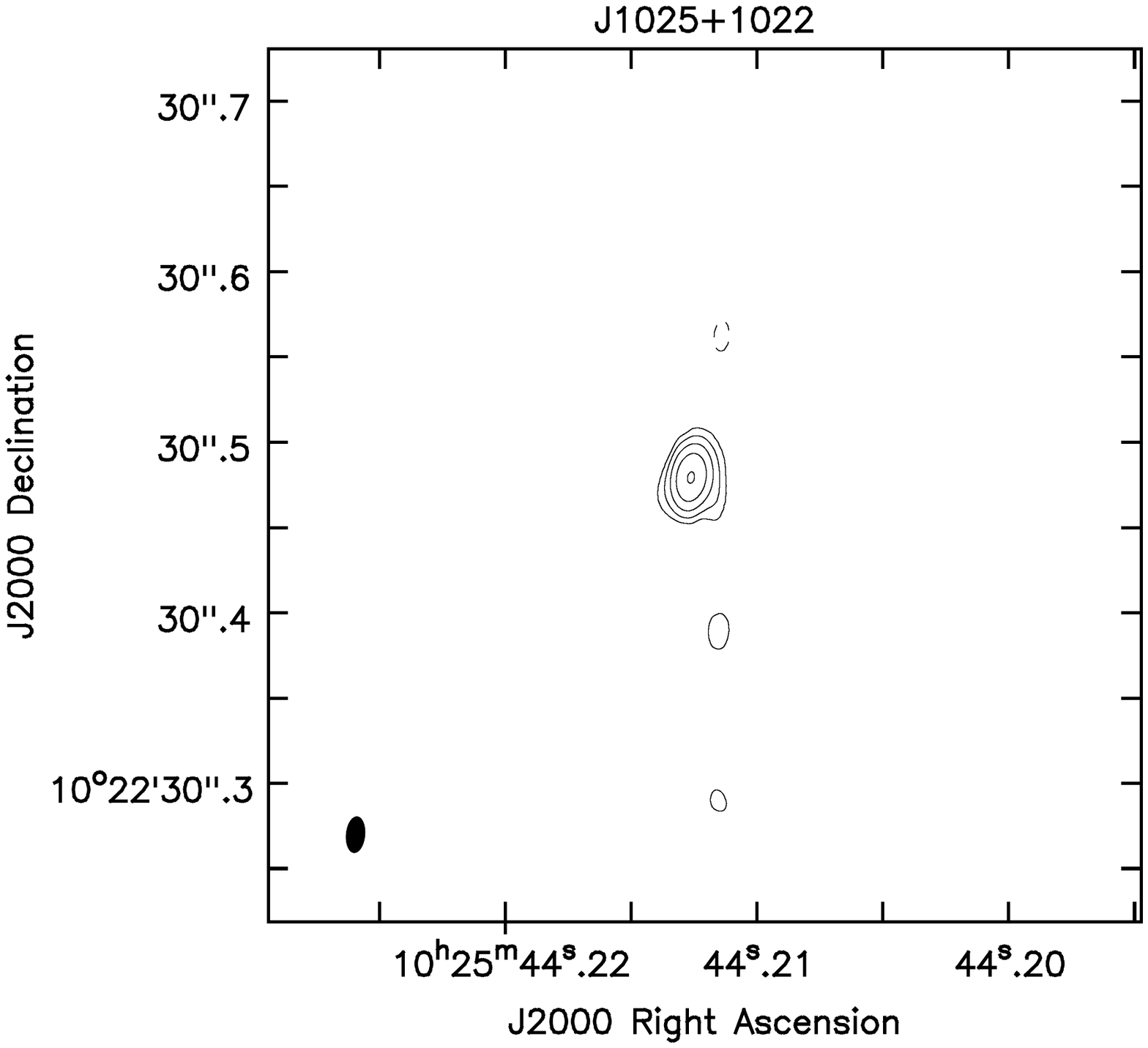}}
\vspace{-2.5cm}
\centerline{
\includegraphics[width=0.3\columnwidth]{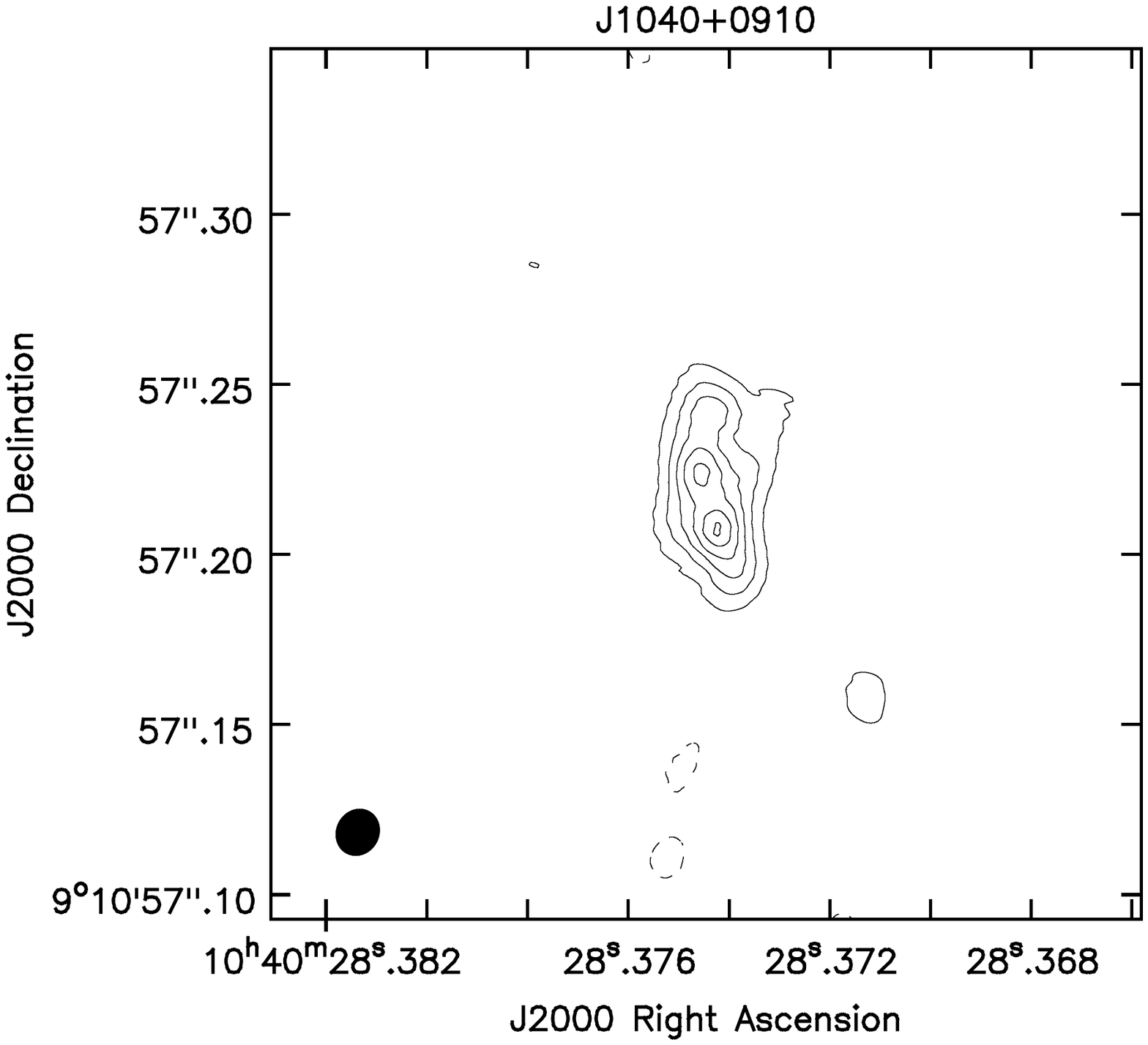}
\hspace{-1cm}
\includegraphics[width=0.3\columnwidth]{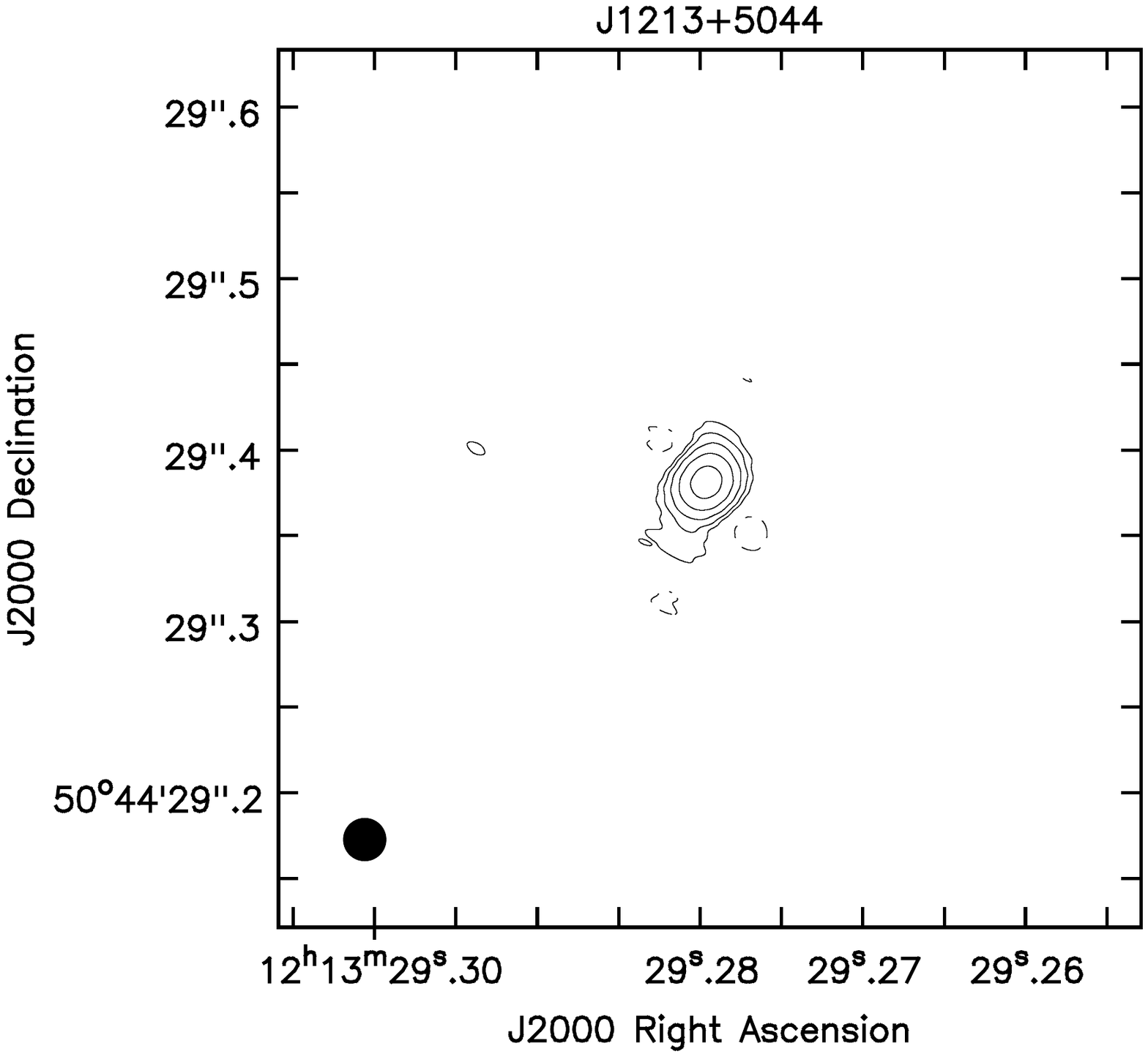}
\hspace{-1cm}
\includegraphics[width=0.3\columnwidth]{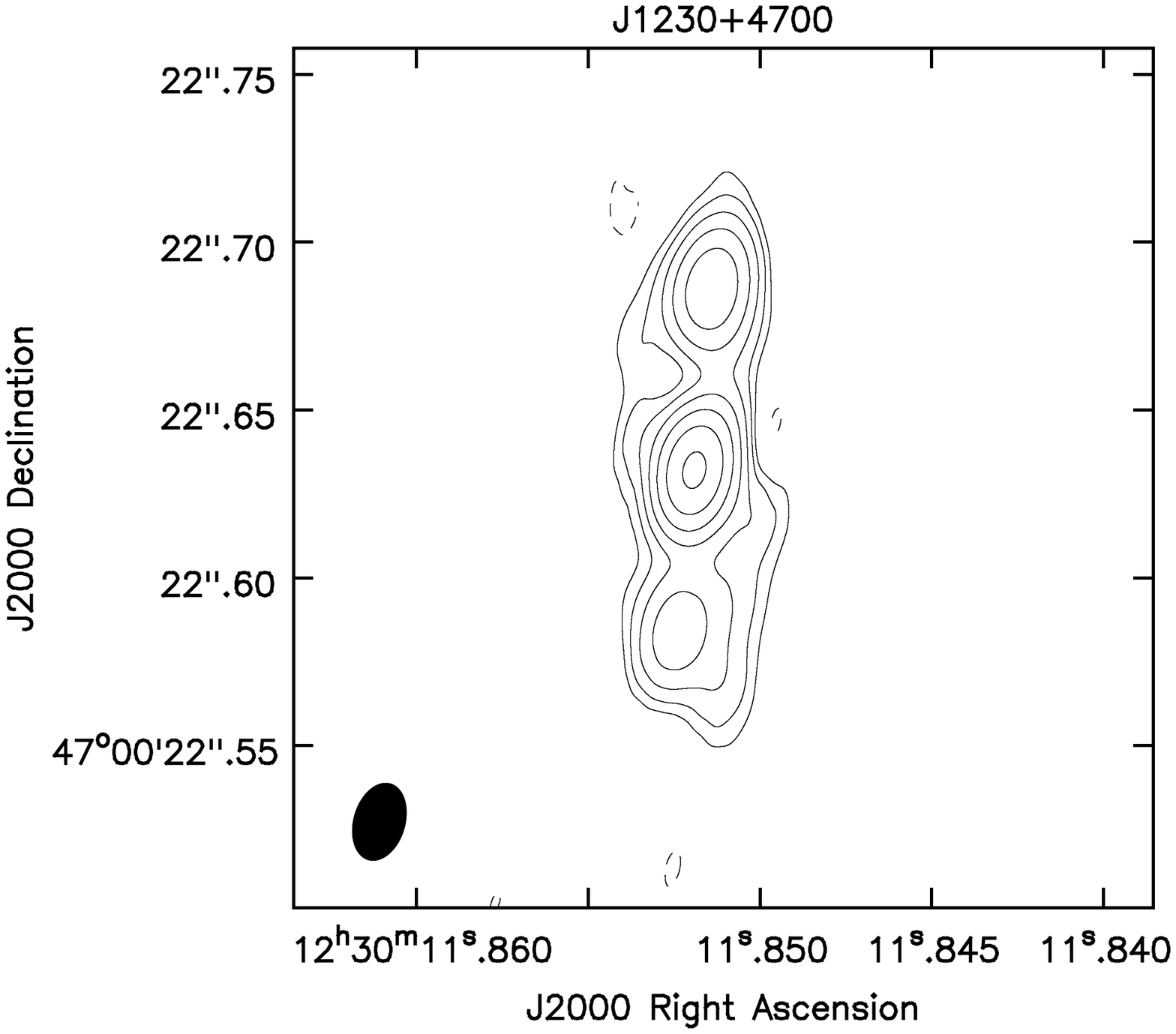}
\hspace{-1cm}
\includegraphics[width=0.3\columnwidth]{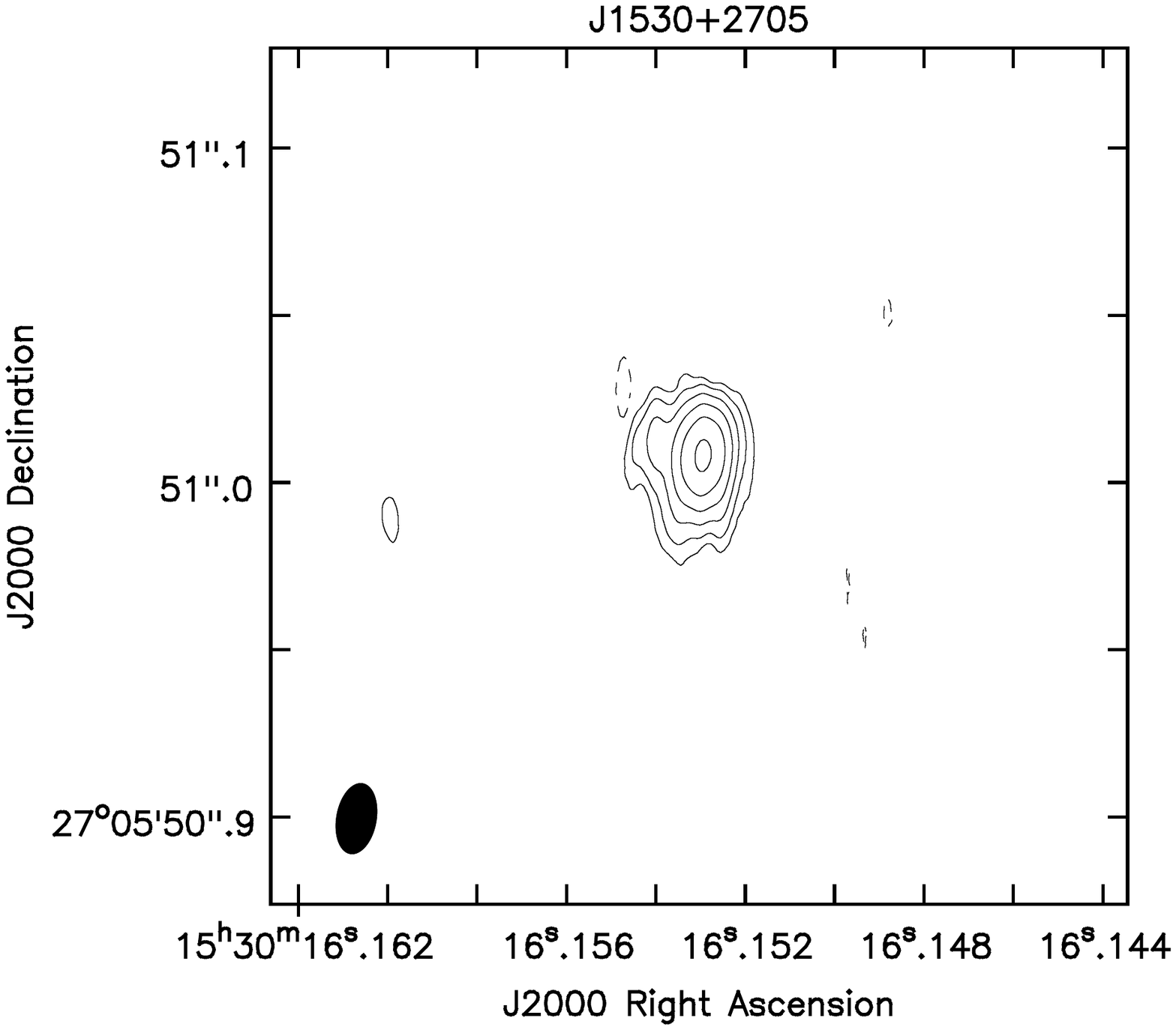}}
\vspace{-2.5cm}
\centerline{
\includegraphics[width=0.3\columnwidth]{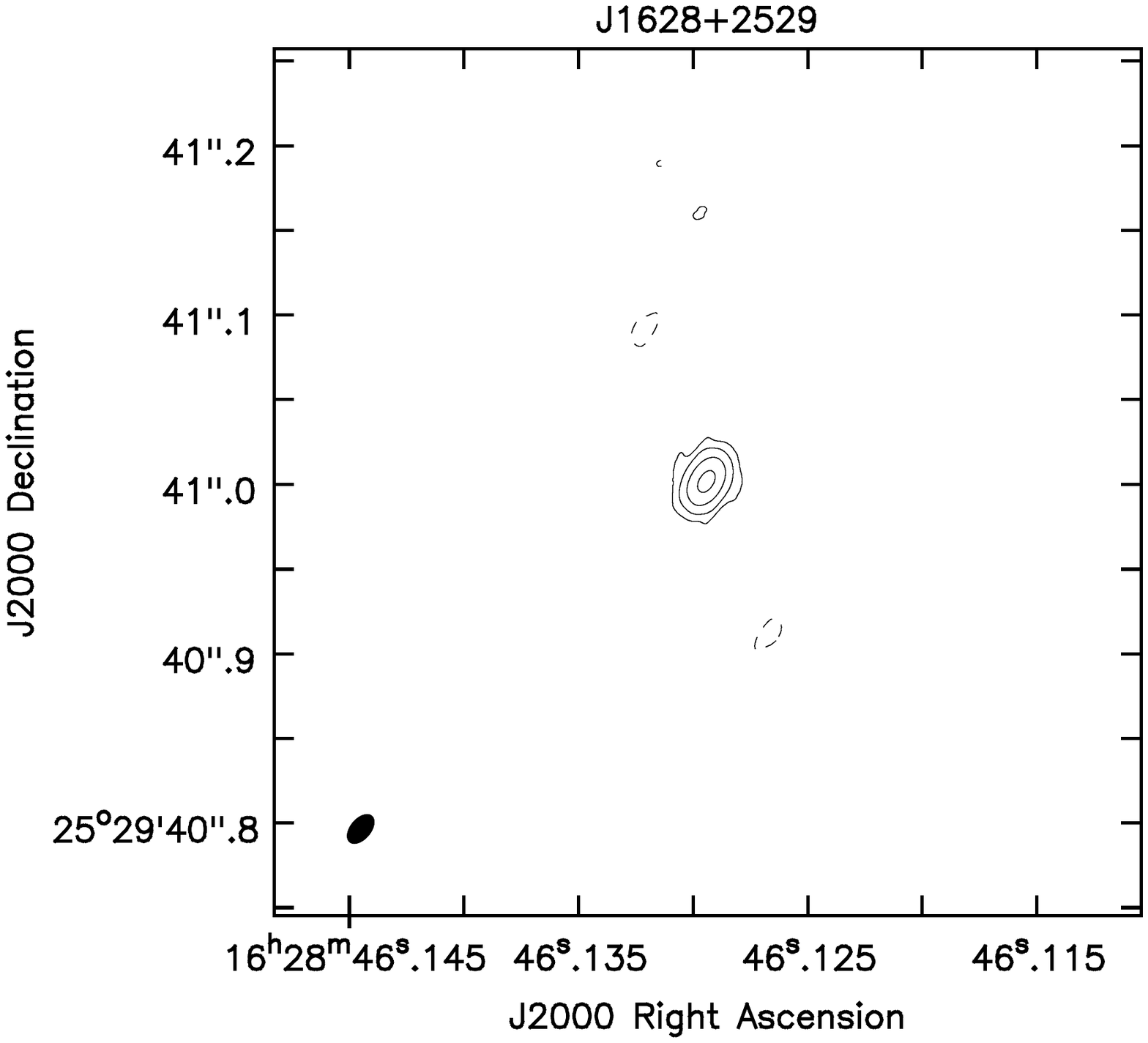}
\hspace{-1cm}
\includegraphics[width=0.3\columnwidth]{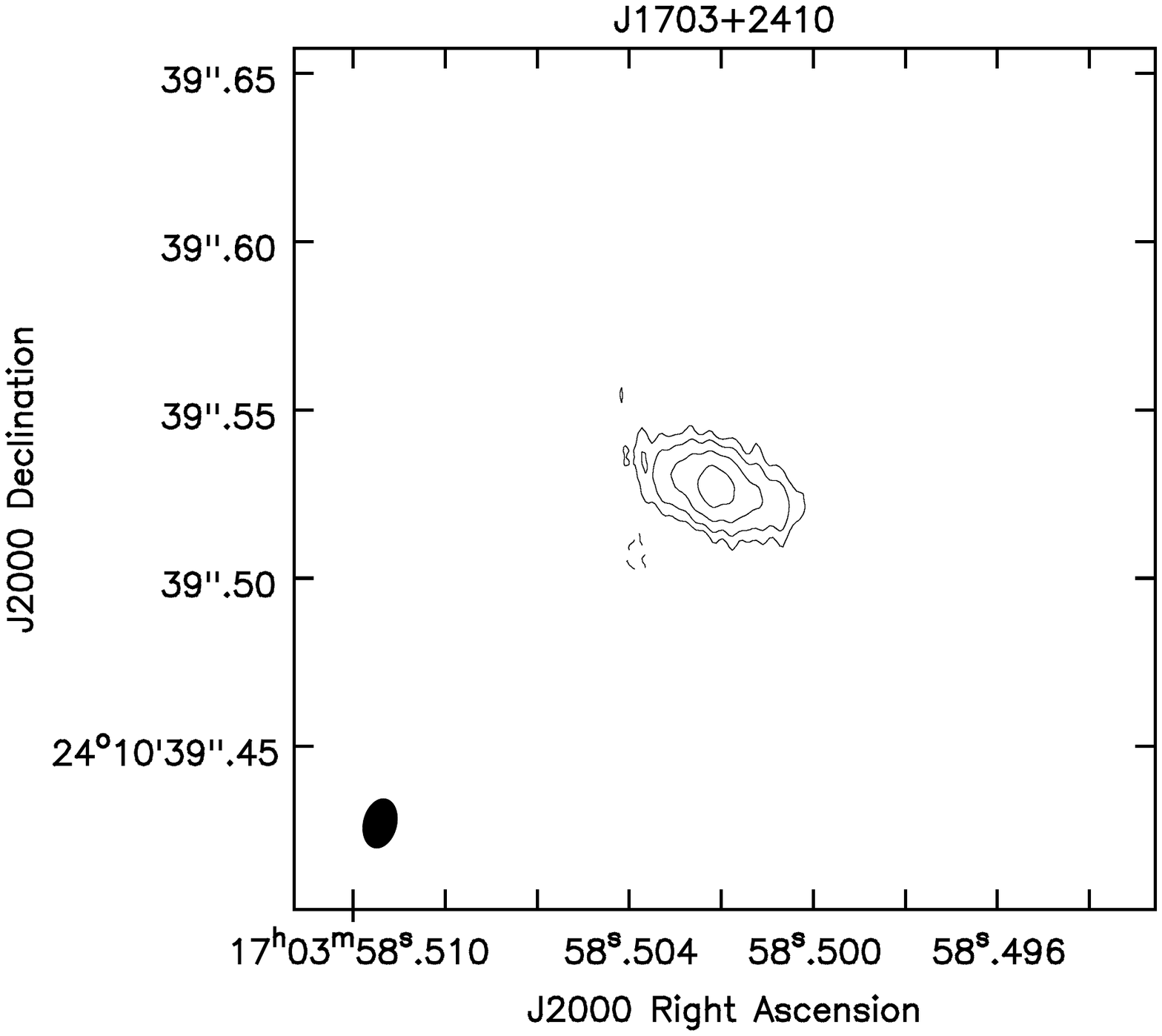}}
\vspace{-2.5cm}
\caption{1.66-GHz EVN observation of 10 FR~0s. The filled area, shown at the bottom-left corner of the images, represents the restoring beam of the maps. Contours and beam parameters are tabled in Tab~\ref{contours}.\label{evn}}
\end{figure}  
\begin{paracol}{2}
\switchcolumn

A catalogue of 104 bona-fide FR~0s galaxies from the SDSS/NVSS sample, named FR0CAT, have been recently set \cite{baldi18a} by including compact sources with: 1) a limit on their deconvolved angular sizes of 4$\arcsec$, corresponding to a linear size $\sim$5 kpc, based on the FIRST images, 2) optical spectrum characteristic of LINERs, 3) red massive ETGs and 4)  z$<$0.05. Eighteen objects were randomly extracted from the FR0CAT sample and observed with the JVLA at L and C band in A-array configuration \cite{baldi19}. To  study the pc-scale structure and then increase the chances to resolve the jets, ten out of these eighteen sources  (see Tab~\ref{tab1}, lower panel) with JVLA flux densities at 1.4 GHz larger than 20 mJy were selected to be observed with the EVN array. These sources have BH masses (10$^{7.5}$ - 10$^{9}$ M$_{\odot}$) and  JVLA radio core luminosities ($\sim$10$^{39}$-10$^{41}$ erg s$^{-1}$) similar to those of the {\it e}MERLIN FR~0 sub-sample. Steep (but flattening at 7.5 GHz), flat and inverted spectra in the range 1.4-4.5 GHz characterise the 10 sources \cite{baldi19}. Three sources reveal kpc-scale emission in the JVLA maps: J0907+3257, similar to J0020-0028, exhibit radio emission from an edge-on disc component; and J1213+5044 and J1703+2410 reveal genuine two-sided jets of a few kpc.

The observations of these 10 objects with EVN (EVN FR~0 sample) at 18 cm (1.66 GHz) occurred in May 2019 in phase-reference mode: target source and a nearby phase calibrator (see list in Tab.~\ref{tab1}) were observed with a $\sim$ 5 minutes cycle. Data were correlated at the JIVE correlator\footnote{https://jive.eu/correlator-overview} in Dwingeloo and the standard EVN pipeline \cite{reynolds02} was used for a first calibration. These data were copied in Bologna where a final calibration and data reduction was done using AIPS. UV-data of calibrators and target sources were examined and bad data were flagged. A standard calibration procedure was followed: a gain correlation solutions were applied to the target sources from the phase and flux calibrators.  A first image was obtained with the 'imagr' task in AIPS and sources with a correlated flux density of a few mJy were self-calibrated to correct the phase only. Analogously to the method explained in Sect~\ref{sect2.1} for the {\it e}MERLIN FR~0 sub-sample, flux densities and sizes for point-like or slightly extended sources are obtained with a Gaussian fit (CASA task 'imfit'), whereas for extended sources the flux density are measured by a flux integration over a region manually selected.

Figure~\ref{evn} depicts the final maps of the 10 FR~0s observed with EVN at 1.66 GHz with a typical angular resolution of 20 mas and rms noise level of several tens of $\mu$Jy (see Tab.\ref{contours}). All the targets are detected with total flux densities of  several mJy (7.5-144 mJy), where the core component appears to be brightest element of the structure (from one fourth to almost the entire flux density of the total emission). In J0907+3257, which exhibit a disc-related extended emission in the JVLA maps \cite{baldi19}, at higher resolution we resolve a triple morphology\footnote{The third south-west component, which is the faintest within the structure, has a peak flux density of 0.45 mJy beam$^{-1}$, thus detected at least at 7$\sigma$ over the rms.}  with a PA, roughly perpendicular to the kpc-scale extended emission. Most of the sources (7/10) appear resolved in elongated structures where additional components are detected in some cases: one-sided jetted (3/7) or double or triple sources (3/7 two-sided). We abstain from the morphological classification of J1040+0910 because it shows a bent triple structure (bending from -12$^{\circ}$ to 24$^{\circ}$), where the core position is ambiguous: in the case of the core as the brightest component of its triple structure, it can be classified as one-sided jetted source, otherwise if the core is the central component, it resembles a two-sided structure. Here we assume that its putative core is the brightest south component. The physical scales of the 10 sources vary between $\sim$5.4 to 133 pc ($\sim$30 mas to 0.15$\arcsec$).

Four of these sources have been already observed with VLBI, VLBA and EVN at 2.3, 5 and 8.4 GHz \cite{cheng18,cheng21} with an angular resolution (up to a factor 10) higher than our observations. J0943+3614 displays a single variable unresolved component with flux densities between 130-270 mJy beam$^{-1}$ with the VLBI \cite{cheng18}, consistent with our EVN detection. While we detect a single core, observations with VLBA and EVN of  J1025+1022 from \cite{cheng21} resolved a two-sided jetted shape with a core component of 60 mJy beam$^{-1}$, a factor $\sim$0.7 lower than our detection. J1213+5044 reveals a VLBA and EVN east-west twin structure of a few pc \cite{cheng21}, included within the (51 pc long) one-sided (PA $\sim$-25) morphology we detect. J1230+4700 has been observed with the VLBI, VLBA and EVN \cite{cheng18,cheng21} and shows a compact North-South two-sided morphology of a few pc, roughly aligned with (128 pc long) triple structure we detect.

EVN resolves the pc-scale core component of several mJy for the all 10 FR~0s with a core dominance  F$_{\rm core}/F_{\rm NVSS}$ ranging on a large interval of values,  0.03--4, indicative of a wide variety (e.g. aligned, variable sources, see Sect.~\ref{discussion}) of radio sources involved in this sample. For the resolved sources, the jet emission consists of 10-60\% of the total emission, values which strongly depend on the morphology (one or two-sided).

\section{Results}
\label{results}

Starting from their sub-arcsec JVLA compact morphologies, the 5-GHz {\it e}MERLIN  and 1.66-GHz EVN observations of 15 FR~0s (at z$<$0.1) have resolved their core component within an extended structure for 11 sources. The central core, which pinpoints the location of the putative BH, represents   the jet base, where we can explore the mechanisms of jet launching and propagation at parsec scale.
For our FR~0 sample, the core flux densities range from sub-mJy to several tens of mJy beam$^{-1}$.
More precisely, the EVN FR~0 sample has higher flux densities than the {\it e}MERLIN sample due to the selection criteria of the EVN observations. The fraction of the pc-scale core flux density with respect to the JVLA component gradually increases with the radio frequencies\footnote{Note that the {\it e}MERLIN  and EVN data are around three years later than the corresponding JVLA observations, respectively, \cite{baldi15} in 2012-2013 and \cite{baldi19} in 2016-2017.}, on average, from around $\sim$10-20\% at 1.4 GHz, to 20-30\% at 4.5 GHz, to 50-70\% at 7.5 GHz. This suggests that a flat-spectrum core emerges from our high-resolution observation, even in the steep-spectrum sources. As exception,  the EVN core flux density of J1530+2705 is higher than those detected previously with the JVLA, a hint for further processes involved (e.g. Doppler boosting, accretion/ejection variability).

The core brightness temperature of the {\it e}MERLIN sample is a few 10$^{4}$ K (Tab.~\ref{emerlin_flux}), lower than the typical value 10$^7$ K, generally expected from relativistic particles emitting synchrotron emission (e.g. from jets or AGN), due to their low  peak flux densities (sub-mJy beam$^{-1}$ level). The much brighter cores and a smaller beam of the EVN sample at 1.66 GHz result in higher brightness temperatures (Tab.~\ref{evn_flux}), i.e. 10$^{7}$-10$^{8}$ K, but  lower than the nominal equipartition brightness temperature $\sim$10$^{10}$ K \cite{readhead94} expected from a relativistic beaming of the jet.

The most remarkable result  of this study is the unprecedented concatenation of the {\it e}MERLIN and JVLA visibilities for five FR~0s.  This procedure turned out to be a positive test to detect sub-kpc scale jets in FR~0s when clearly resolved jetted structure both at arcsec and mas scale resolution do not emerge.  Four out of five FR~0s reveal the presence of two-sided jets with radio extents $\sim$0.3--1.5 kpc.

\begin{figure}[h]
\hspace{-2cm}
\includegraphics[width=0.8\textwidth,angle=180]{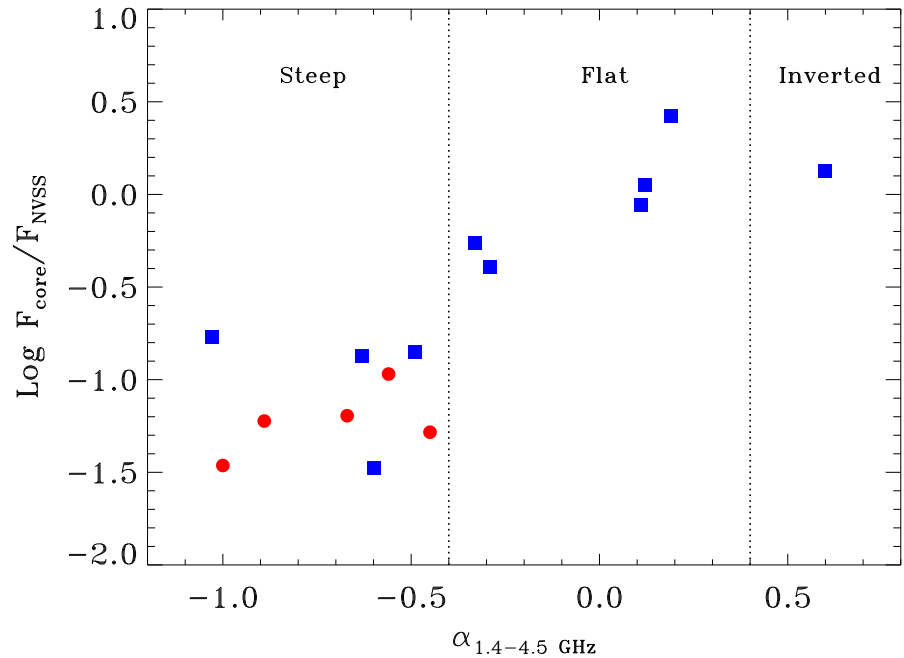}
\vspace{-1cm}
\caption{The 1.4--4.5 GHz spectral slope ($S_\nu \propto \nu^{\alpha}$) measured from the JVLA data \cite{baldi15,baldi19} as function of the core dominance, estimated as ratio of the flux density of the pc-scale core component measured from our {\it e}MERLIN and EVN data and the total extended emission from NVSS flux density. Red circles mark the {\it e}MERLIN FR~0s and blue squares the EVN FR~0s. The vertical dashed lines denotes the steep-spectrum sources ($\alpha$ $<$ -0.4), the flat-spectrum sources (-0.4 $<$ $\alpha$ $<$ 0.4) and the inverted-spectrum sources ($\alpha >$ 0.4).\label{slope_cd}}
\end{figure}

Since FR~0s have been defined as compact radio sources which lack substantial extended emission, the detection of jetted structures for most of the sources from tens of pc to kpc is surprising.  The pc-scale jetted shapes are typically aligned with the JVLA jets or those detected at higher resolution  by \cite{cheng18,cheng21}. The radio morphologies of the pc-scale extended jets are predominantly two sided (two-sided:one-sided=7:3) showing lobes and plumes\footnote{J1040+0910 has been excluded because of  the precarious identification of core.}. Among the two-sided jet group, core brightened morphologies are more frequent (core-brightened:edge-brightened=4:3) than the edge-brightened morphologies (triple and lobed), despite of an evident difficulty of classification due to the low brightness of the jets. 

Particularly interesting are the two cases of RGs which emit kpc-scale disc emission probably from  edge-on lenticular galaxies. At higher resolution, J0907+3257 reveals the presence of a triple structure of 133 pc with PA roughly perpendicular to the extended disc emission, whereas J0020-0028 shows scattered radio emission along the JVLA radio structure with a north-west marginally elongated core towards the JVLA-detected twin shape.

Considering the radio spectra of the sample, we note a relation between the core dominance of the source (i.e ratio between the pc-scale core measured from our maps and the total radio emission measured from NVSS) and the spectral slope  ($S_\nu \propto \nu^{\alpha}$) measured between 1.4 and 4.5 GHz from JVLA data taken from \cite{baldi15,baldi19} (see Fig.~\ref{slope_cd}). The contribution from the core component of the total extended emission increases with the spectral flattening. This is in line with a  tendency of the flat/inverted-spectrum sources of being unresolved or one-sided jetted. In addition, it is noteworthy to point out that three sources (J0943+3614, J1025+1022, J1530+2705), which are characterised by an inverted/flat spectrum and by a substantial flux increase of their JVLA detections with respect to their NVSS-FIRST measurements by a factor 1.6-3.2 \cite{baldi15,baldi19},  show a (one-sided) compact core, possibly indicating of a marginal alignment of the source with the line of sight.

\begin{figure}[h]
\hspace{-2cm}
\includegraphics[width=0.9\textwidth,angle=180]{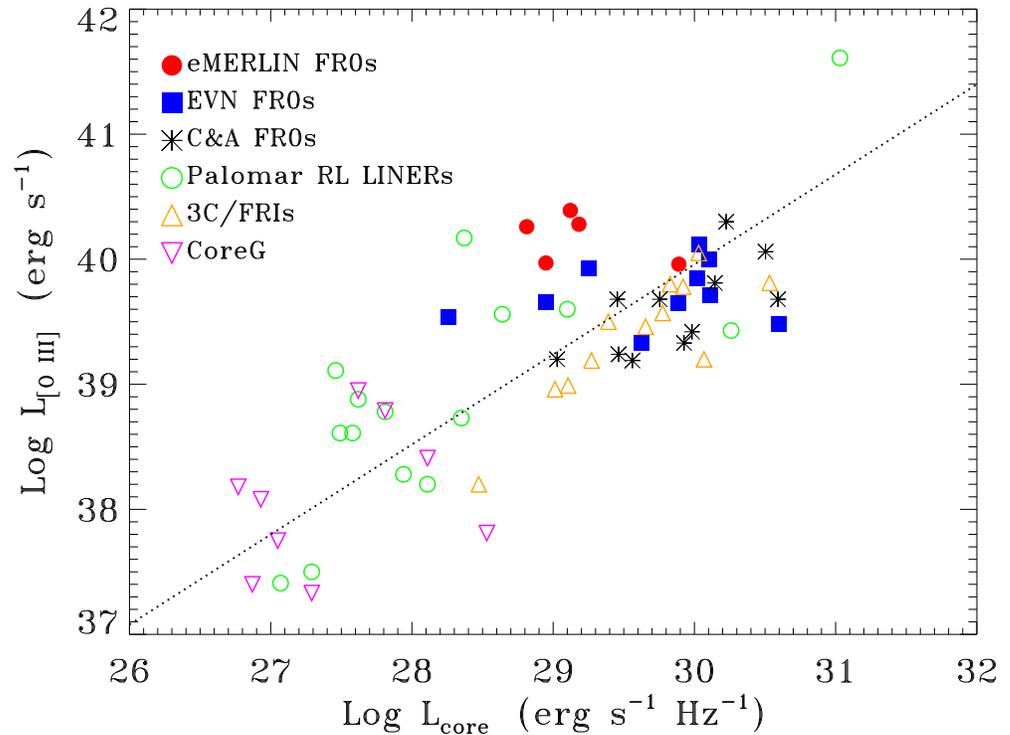}
\vspace{-1cm}
\caption{Parsec-scale core radio power (erg s$^{-1}$ Hz$^{-1}$) vs. [O III] line luminosity (erg s$^{-1}$) for different samples of LINER-type RL AGN hosted in ETGs with core-brightened morphologies (see the legend): red dots for the {\it e}MERLIN FR~0 sub-sample, blue filled squares for the EVN FR~0 sub-sample, black stars for FR~0s from \cite{cheng18,cheng21}, green empty circle for RL Palomar LINERs, upwards orange triangles for 3C/FR~Is, downward pink triangle for Core Galaxies (see Sect.~\ref{results} for details). The dotted
line indicates the best linear correlation found for all the sources.\label{lcorelo3}}
\end{figure}

The pc-scale (peak) core luminosities L$_{\rm core}$ of the {\it e}MERLIN and EVN FR~0 sub-samples range between 10$^{21.3}$ - 10$^{23.6}$ W Hz$^{-1}$ (integrated luminosities $\nu$L$_{\rm core}$ $\sim$ 10$^{37.5}$ - 10$^{39.8}$ erg s$^{-1}$), a factor 10-100 weaker that the FR~0s studied by \cite{cheng18,cheng21} (10$^{23}$-10$^{24}$ W Hz$^{-1}$). The pc-scale cores detected by {\it e}MERLIN and EVN offer a more robust estimate of the jet base energetics than the sub-arcsec cores detected by JVLA. In this interpretation, Figure~\ref{lcorelo3} represents a diagnostic plot, radio vs [O~III] luminosity, i.e. a proxy of the jet energetics vs a proxy of the AGN bolometric luminosities\footnote{[O~III] line luminosity is a good indicator of the bolometric AGN luminosity ($L_{\rm Bol}$ = 3500$\times L_{\rm [O~III]}$, \cite{heckman04,heckman14}) for low-luminosity AGN.}. The two quantities are expected to correlate following the radio-line relationship valid for FR~0s and FR~Is \cite{baldi15,baldi19}, suggesting that  radio emission efficiency, i.e. the fraction of the radio emission produced with respect to
the AGN accretion power, is constant between FR~0s and FR~Is. As we can note from Fig.~\ref{lcorelo3}, the [O~III] luminosities, L$_{\rm [O~III]}$, of the FR~0s studied in this work, $\sim$10$^{39.3}$ - 10$^{40.4}$ erg s$^{-1}$, spread on a narrower range  by a factor 10 than the corresponding core luminosities. This discrepancy from a 1-to-1 correspondence between the core and line distributions could be the result of a sharper dependence on possible Doppler flux boosting or larger core variability on parsec scales than those from a sub-arcsec view of the JVLA cores, apart from a possible flattening of the slope of the radio-line correlation.  If we include the FR~0s studied  with the VLBI technique by \cite{cheng18,cheng21} (black stars in Fig.~\ref{lcorelo3}), the data points cluster in limited intervals of radio and [O~III] luminosities\footnote{This is because the FR~0 sample is selected at low luminosities, reaching the
low-end limit of the flux-limited SDSS survey, L$_{\rm [O~III]} \sim$10$^{39}$ erg s$^{-1}$ for galaxies at z$<$0.1.}, preventing from a statistical study of the presence of a L$_{\rm core}$-L$_{\rm [O~III]}$ correlation valid for FR~0s.

\begin{figure}[h]
\hspace{-1cm}
\includegraphics[width=0.8\textwidth,angle=180]{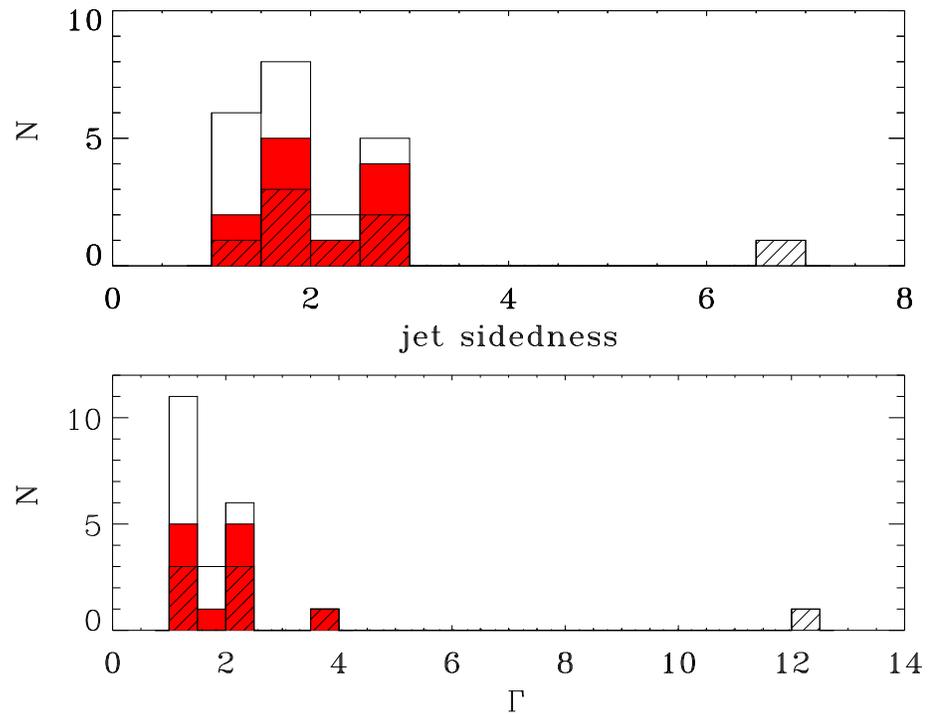}
\vspace{-1cm}
\caption{Upper panel: histogram of jet-to-counter-jet sidedness from our sample (white histogram) and the sample from \cite{cheng18,cheng21} (red histogram) obtained from VLBI, EVN and VLBA data. Lower panel: distribution of the jet bulk Lorentz factor $\Gamma$, estimated from the jet sidedness (see text in Sect.~\ref{results}). The shaded areas represent the lower limits of the jet sidedness and $\Gamma$ for the one-sided sources.  \label{jetsd}}
\end{figure}  

Since an affinity between the nuclear properties of FR~0s and FR~Is  has been clearly assessed \cite{baldi15,baldi19,torresi18} and low-power RGs have been found to be the scaled-down version of FR~Is (e.g., \cite{nagar05,balmaverde06a,balmaverde06b,baldi18b,baldi21b}), it is plausible to search for a general connection between the jet and accretion strength of a population of low and high power RGs with comparable properties: hosted in massive ETGs and characterised by a LINER spectrum\footnote{We exclude the FR~II LINERs, because  the nature of their central engine and its relation with the FR~I LINERs are still not fully understood \cite{heckman14,tadhunter16,macconi20}.}. With the purpose of  building such a sample of local (z$<$0.05) LINER-type RL AGN, we have collected VLBI, VLBA, EVN and {\it e}MERLIN data at 1.4 and 5 GHz available from the literature ([O~III] line emission from \cite{ho97,buttiglione10}), see Fig~\ref{lcorelo3}, for: LINERs from the Palomar sample of nearby galaxies \cite{ho95} (green circles, from \cite{anderson05,nagar05,krips07,liuzzo09}), which appear as unresolved cores or core-jet structures at JVLA scales in the radio band and are classified as RL by \cite{baldi21b}; 3C/FR~Is (all LINERs, open upwards triangles, from \cite{xu99,jones97,giovannini05,liuzzo09}); and Core Galaxies (open downward triangles, from \cite{nagar05}), which have LINER spectra and are known to be miniature FR~Is \cite{balmaverde06a}. With the inclusion of these RL AGN in the L$_{\rm core}$-L$_{\rm [O~III]}$ plot, it is evident that FR~0 data points belong to a broad sequence between the two luminosities correlated on $\sim$4 order of magnitudes. We fit the data points present in this sequence with a linear (in a $\log$-$\log$ plot, hereafter) relation and we find a correlation in the form L$_{\rm [\rm O~III]}$ $\propto$  L$_{\rm core}^{0.72\pm0.10}$ with a Pearson correlation coefficient (r-value) of 0.782 which indicates that the two quantities do not  correlate with a probability of 3$\times$10$^{-11}$. This statistically-robust relation corroborates the idea that nuclei of FR~0s are similar to those of LINER-like RL AGN. The large scatter of the correlation, $\sim$0.27 dex, could be caused by the Doppler boosting, the nuclear variability and the different data frequencies (particularly sensitive in the case of steep spectrum). Furthermore, the slope of the established linear correlation is consistent with that found for nearby low-luminosity RL LINERs studied with {\it e}MERLIN \cite{baldi21b}. However, a study on a large sample of FR~0s at lower radio luminosities is still needed to confirm the existence of this linear regression.

With the purpose of studying the ejection mechanism in FR~0s, for the extended sources of our {\it e}MERLIN and EVN sub-samples, we estimate the jet/counter-jet ratio measuring the brightness ratio in two symmetric regions as near as possible to the nuclear emission by avoiding the core and considering the surface brightness at similar distance from the core. The jet-to-counter-jet sidedness ratios measured for the JVLA+{\it e}MERLIN and EVN sub-samples (Tab~\ref{emerlin_flux} and \ref{evn_flux}) are between 1.1-1.8 and 1-3, respectively. For each source also resolved by the JVLA, the mas-scale jet ratio is higher than the corresponding value measured with JVLA at sub-arcsec scale (in the range, 1-2 \cite{baldi19}). Figure~\ref{jetsd} (upper panel) displays the pc-scale jet sidedness distribution measured from our data and from VLBI observations taken from \cite{cheng18,cheng21}\footnote{We measure the jet components for the one-sided and two-sided jets at L and C band (or X band if the previous two bands are not available) from VLBI, VLBA and EVN data from \cite{cheng18,cheng21}.}. The distribution of jet-to-counter-jet brightness ratio for our FR~0s and the sample from \cite{cheng18} goes between 1 and 3 with one exceptional value at 6.5 (J0907+3257). The shaded histogram in Fig.~\ref{jetsd} (upper panel) represents the lower limits for the one-sided jetted sources.

From this parameter we estimate the bulk Lorentz $\Gamma$ factor of the jet following the procedure discussed by \cite{bassi18}. Since the jet parameters are unknown, this method assumes the value for angle $\theta_m$ that maximizes $\beta$, i.e. cos $\theta_m$ = $\beta$, where $\beta = v/c$ is the intrinsic jet speed, in units of $c$ and $\theta$ is the angle of the jet to the line of sight. With this simplification, the maximum $\Gamma_{\rm bulk}$ is
\end{paracol}
\nointerlineskip
\begin{equation}
\Gamma_{\rm bulk} = \biggl({R^{m} + 1  \over 2} \biggr)^{0.5} 
\end{equation}
\begin{paracol}{2}
\switchcolumn

where R is the jet sidedness and $m =  k - \alpha$, where $k$ is the parameter that
accounts for the geometry of the ejecta, with k = 2 for a continuous jet and $\alpha$ is the spectral index of the emission ($S_\nu \propto \nu^{\alpha}$, taken from Tab.~\ref{emerlin_flux} and Tab.~\ref{evn_flux} and from \cite{cheng18,cheng21}). With these assumptions, the range of $\Gamma_{\rm bulk}$ measured for our sample and the FR~0s from \cite{cheng18,cheng21} is between $\sim$1 and 2.5 with two values at $\sim$3.5 and 12 for the targets  J0933+1009 and  J0907+3257, respectively (Fig.\ref{jetsd}, lower panel). We plan to re-observe these last two FR~0s with VLBI array to check their particular extreme jet properties with respect to the rest of the sample.

\section{Discussion}
\label{discussion}

Our {\it e}MERLIN and EVN observations of 15 FR~0s confirm the ability of this class of RGs to launch pc-scale jets, at lower luminosities ($\lesssim$10$^{23}$ W Hz$^{-1}$) than previous VLBI studies of FR~0s. Radio structures with extents from a few tens to hundreds of pc are present for $\sim$86 \% of the FR~0s studied so far with VLBI technique (this work, \cite{cheng18,cheng21}), despite 4 FR~0s reveal their jetted structures only after combining JVLA and {\it e}MERLIN datasets. This result eventually attests the idea that FR~0s are genuine RGs with extended jets, although of limited size. The resolved jetted structures typically host a steep-spectrum core, while the flat/inverted spectrum FR~0s appear as one-sided jets or individual core components. This morphology-spectrum connection is the result of a core dominance with respect to the total emission, increasing with the flattening/inverting of the radio spectra. This highlights the power of the VLBI-based observations in resolving the pc-scale flat-spectrum core in 'compact' sources like FR~0s.

The effective capability of launching pc-scale of FR~0 jets adds up to a further similarity with classical FR~Is, which generally exhibit core-brightened\footnote{However, it is correct to point out that the observed pc-scale core-brightening morphology of FR~Is could be the effect of a selection bias because only FR~Is with bright cores have been observed with the VLBI \cite{liuzzo09}.} radio morphologies at parsec scale \cite{fanti87,venturi95,giovannini05}. At kpc-scale, about half of the FR Is shows large jet asymmetries with a jet/counter-jet
flux ratio larger than 2 \cite{bridle84,parma87}. At smaller scales, the jet-to-counter-jet sidedness ratio of FR~Is \cite{giovannini90,venturi95,giovannini98,giovannini01} is generally larger than that measured with the JVLA \cite{xu00}. This morphological jet variation of FR~Is at different scales is interpreted with a change of the jet bulk speed, initially from relativistic, $\Gamma>$3, to sub-relativistic speeds on kpc scales by decelerating for entrainment \cite{bicknell84,kharb12b}. In comparison to FR~Is, for the sample of 22 FR~0s studied at parsec scale (our work, \cite{cheng18,cheng21}), the jet-to-counter-jet ratio is less prominent: about one third of FR~0s has jet sidedness larger than 2. With inevitable assumptions on jet properties, the estimated FR~0 jet bulk $\Gamma$ factor, maximised to reproduce the observed distribution of jet sidedness, ranges between 1-2.5 for $\sim$90 \% of the FR~0 sample. This is in agreement with results from \cite{cheng18,cheng21} that FR~0 jets are mildly relativistic with low jet proper motions. However, a  proper systematic analysis on larger samples of FR~0s matching in redshifts and luminosities with FR~Is is needed to draw a final conclusion on the jet bulk speed difference between the two populations.

Another outstanding result of this study, which strengthens the connection between FR~0s and FR~Is, is the presence of a broad  correlation between the pc-scale core luminosities and the [O~III] line luminosities, used as proxy of the AGN strength, established for low-power LINER-like RGs, which encompass FR~0s, FR~Is, and low-luminosity RL LINERs in general.  This result suggests that RL LINERs, other than sharing the same type of host (massive ETGs with BH mass $>10^{7.5}$ M$_{\odot}$), also shares a single type of central engine, able to launch (from low to high power) core-dominated jets \cite{capetti17,baldi18b,baldi21a} with (from mildly to highly) relativistic speeds. A radiatively inefficient accretion disc is the most plausible engine to account for the low Eddington ratios and low bolometric luminosities of FR~0s and FR~Is \cite{balmaverde06b,balmaverde08,torresi18}. This hypothesis is supported by the idea of an ADAF disc coupled with a jet, typically attributed to the RL LINER population in general, in agreement with previous studies (e.g. \cite{falcke04,chiaberge05,nemmen14,baldi21b}). In fact, RIAF discs are supposed to be efficient at emanating jets, even at low energies, as suggested by theoretical studies including analytical works (e.g., \cite{narayan94,meier01,nemmen07} and numerical simulations (e.g., \cite{tchekhovskoy11,mckinney12}).

The best scenario, which would simultaneously predict low-brightness pc-scale jets of FR~0s and a disc-jet coupling and host conditions similar to those of more powerful FR~Is, but  differing in large-scale environment, is a specific cosmological evolution of RGs based on BH spin and clustering. In an accretion-driven scenario, FR~0s host prograde low-spinning BHs, which will eventually evolve into highly spinning  FR~Is when  the accreting material reaches $\sim$30\% of the initial BH mass \cite{garofalo19}. The poorer environment of FR~0s, as compared to the one of FR~Is, would favor the longer stage of FR~0s due to the limited gas availability and thus would account for the larger space density of compact RGs than the extended ones \cite{capetti20b}. Assuming a BH spin-$\Gamma_{\rm jet}$ dependence (e.g., \cite{tchekhovskoy10,maraschi12}), the lower bulk Lorentz factor of FR~0s would be caused by their lower BH spin, limiting so their extracted energy to launch mildly-relativistic jets. Instead, in a merger-driven scenario, assuming that the BH spin is the main result of a BH-BH coalescence event, in a poorer environment major mergers of equal mass galaxies are less frequent \cite{stewart09,jian12}, resulting in low probabilities of reproducing highly-spinning BHs, a condition required to create extended relativistic jets. Therefore, the large-scale environment generates different merger histories and different BH spin distributions between FR~0s and FR~Is, resulting eventually in a large and small population of RGs with compact and fully-fledged morphologies, respectively.

While the majority of FR~0 population conforms with the idea that
they are compact core-dominated sources with mildly-relativistic core-brightened jets, a small fraction exhibits different radio features from this picture. In fact, different studies at low and high radio frequencies suggested that FR~0s  certainly are a mixed population of low-power radio sources which includes genuine young RGs, such as CSS and GPS \cite{sadler14,odea21,mihahilov21b}. In our sample, we must mention that  one source (J0943+3614) has an inverted spectrum, similar to what is typically seen in GHz-peaked sources \cite{odea21}, but previous VLBI and EVN observations of this target resolve only a core component. In addition, two sources (J0907+3257 and J1230+4700) show a mas-scale double-lobed structures consistent with a classical picture of a new-born RG. The three variable radio sources (J0943+3614, J1025+1022, J1530+2705), instead, could reconcile with the picture of the aligned counterparts of FR~0s, as the BL\,Lacs population has been claimed as the parental population of FR~0s \cite{massaro20}.  All these possible exceptions to the general FR~0 rule contribute to an approximately estimated $\sim$10\% of contamination present in the FR~0 population \cite{baldi18a}, where young RGs, blazars, radio-quiet AGN and low-luminosity compact sources\footnote{Low-luminosity compact objects have been recently addressed by \cite{kunert09}, as kpc-scale radio sources with possibly fading steep-spectrum disrupted jets and short-lived activity \cite{kunert14}.} \cite{baldi15,sadler14,odea21,mihahilov21b} would erroneously fall in the FR~0 category, if their radio-optical properties are not deeply analysed. Nevertheless, it is possible that a genuine population of FR~I progenitors or aligned weak BL\,Lacs whose properties have a large overlap with those of misaligned FR~0s, cannot be fully disentangled from the bona-fide FR~0 population.

\section{Summary and Conclusions}
\label{s&c}

We present {\it e}MERLIN and EVN observations of a sample of 15 FR~0s, respectively, 5 sources at 5 GHz and 10 sources at 1.66 GHz. Parsec scale one- and two-sided jets  have been detected for 11 sources of the sample and cores for the remaining objects with angular resolution of a few mas and sensitivity of several tens of $\mu$Jy. 

Since we also own calibrated JVLA C-band data for the five FR~0s observed with {\it e}MERLIN, for the first time we combined the visibility datasets of the two arrays in this band for this population of apparently jet-less compact RGs.  This method aims at probing  the intermediate scales of the jet formation and propagation between the kpc-scale emission lacking in FR~0s and the sub-kpc core resolved with JVLA. This procedure turns out to be successful in detecting pc-scale jets, which were missing in the two original datasets. This result means that the jets appear unresolved and resolved out, respectively, in the JVLA and {\it e}MERLIN maps. We can thus conclude that FR~0s, although apparently lacking of extended emission, are effectively able to emanate pc-scale jets, whose low brightness make them hard to isolate and detect. The FR~0s, which do not show extended jets, are typically variable and/or one-sided or having unresolved core, suggesting a possible jet orientation effect. In any case, the combination of long and short baseline represents a powerful tool to study the jet properties of the FR~0 population. 

The compact component detected with mas-scale angular resolution, at the centre of  the resolved jetted structures of FR~0s studied with the VLBI technique, delimits the parsec-scale section of the jet along its extension. This radio-emitting region of a few pc represents a closer look to the jet launching region and thus a better proxy of the jet energetics than previous radio studies. The core identification of FR~0s enabled us to carry on an unbiased comparison with the rest of widely-studied population of RGs. In fact, it has been found that FR~0s, together with a homogeneous population of low-power RGs, all characterised by a LINER optical spectrum, an early-type host and large BH masses ($>$10$^{7.5}$ M$_{\odot}$, satisfying the radio-loudness criteria \cite{chiaberge11}) appear to follow a broad correlation between the pc-scale core luminosities, considered as good proxy of the jet kinetic power, and the [O~III] line luminosities, indicative of the accretion power. This result sets a disc-jet connection similar to FR~Is, where a RIAF disc supports the launch of a core-brightened radio structure, valid for FR~0s and generally for low-power LINER-type RGs. In conclusion, a particular set of nuclear and host properties defines a homogeneous class of RGs with a common accretion-ejection mode, from low-luminosity RL AGN to powerful FR~Is. However, although they show large differences in radio behavior from the point of view of their sizes, luminosities, and morphologies, LINER-like low-luminosity AGN and FR~0s represent the low-end part of this continuous population distribution, which ends with fully-fledged FR~Is. An increasing radio luminosity, possibly due to higher BH spins or other higher values of the internal parameters (BH mass, magnetic field, etc.), would smoothly connect the low-power RGs and FR~0s to the FR~Is. 

Low and high frequency radio, optical and X-ray observations of FR~0s have shown that FR~0s differ from FR~Is for the large space density, the large-scale radio structure and Mpc-scale environment,  unlike their nuclear and host properties which appear indistinguishable. These similarities and differences have been interpreted in the framework of RG evolution: FR~0 state consists of a long phase of the life-time of RGs, which inhabit moderately poor environment and are powered by low spinning BHs, which then struggle to spin up for the low accretion rate. This scenario predicts that FR~0s are characterised by low jet bulk speed, assuming a linear connection between the relativistic flow of the jets and the BH spin, from where the jet energy is extracted. This interpretation is supported by our {\it e}MERLIN and EVN study: the bulk jet speeds of FR~0s, estimated from the pc-scale sidedness, are mostly in the range 1-2.5, lower than typical values $>$3 found for FR~Is. In a two-zone jet model with an inner fast spine and a slower layer \cite{ghisellini05}, a weak/short (mildly) relativistic spine, which then possibly decelerates and disrupts within the galaxy via gas entrainment \cite{bicknell84,laing96,laing14}, could reconcile with the high-energy and radio properties of FR~0s \cite{baldi19b}. Nevertheless, more dedicated studies on a statistically-complete sample of FR~0s with VLBI-technique observations are needed to confirm their slower jets than those of FR~Is and to eventually understand the physical mechanism of jet launching/propagation in this still-marginally explored class of compact RGs.

The fast approaching high-resolution and high-sensitivity radio facilities, such as the SKA \cite{kapinska15} and the ngVLA \cite{nyland18}, offer promising opportunities to place more stringent constraints on the jet speed and unearth large population of FR~0s in the local Universe. The observations with these up-coming observatories will shed light on the mechanisms of accretion-ejection coupling in this abundant population of compact RGs within a general picture of evolutionary and orientation-dependent RL AGN unification scheme.

\vspace{6pt} 



\authorcontributions{R.D.~Baldi had analysed the {\it e}MERLIN data and combined the {\it e}MERLIN and JVLA data. G. Giovannini has analysed the EVN dataset. All the authors contribute to the discussion of the result of this work. R.D.~Baldi has written the article and the co-authors gave comments and suggestions to improve the quality of the manuscript.}

\funding{This research received no external funding.}

\acknowledgments{We thank the referees for improving the quality of the paper. {\it e}MERLIN is a National Facility operated by the University of Manchester at Jodrell Bank Observatory on behalf of STFC, part of UK Research and Innovation. The European VLBI Network is a joint facility of independent European, African, Asian, and North American radio astronomy institutes. Scientific results from data presented in this publication are derived from the following EVN project code: E17C004.}

\conflictsofinterest{The authors declare no conflict of interest.} 



\abbreviations{The following abbreviations are used in this manuscript:\\

\noindent 
\begin{tabular}{@{}ll}
RIAF \,\,\,\ Radiatively Inefficient Accretion Flow\\
RFI \,\,\,\ radio-frequency interference\\
AIPS \,\,\,\ Astronomical Image Processing System\\
eMERLIN \,\,\,\ {\it e}-Multi-Element Radio Linked Interferometer Network\\
SDSS  \,\,\,\ Sloan Digital Sky Survey \\
FIRST  \,\,\,\ Faint Images of the Radio Sky at Twenty centimeters survey \\
NVSS   \,\,\,\ National Radio Astronomy Observatory Very Large Array Sky Survey \\
JVLA \,\,\,\ Jansky Very Large Array\\
LOFAR \,\,\,\ Low Frequency Array \\
TGSS \,\,\,\ TIFR GMRT Sky Survey \\
SED \,\,\,\ Spectral Energy Distribution \\
VLBI \,\,\,\ Very-long-baseline interferometry \\
VLBA \,\,\,\ Very Long Baseline Array \\
EVN  \,\,\,\ European VLBI Network \\
CSS \,\,\,\ Compact Steep spectrum\\
GPS \,\,\,\ GHz-peaked sources\\
\end{tabular}}

\appendixtitles{no} 
\appendixstart
\appendix
\section{}
\label{app}

We present Table~\ref{contours}, which provides the radio contours, rms
and the properties of the restoring beams of the {\it e}MERLIN, combined JVLA-{\it e}MERLIN and EVN maps of our sample.

\begin{specialtable}
\caption{Column description: 1) source name; 
(2) FWHM of the elliptical Gaussian restoring beam (in arcsec$\times$arcsec) of the maps from {\it e}MERLIN (Fig.~\ref{emerlin}), combined {\it e}MERLIN and JVLA  (Fig.~\ref{combined}) and EVN (Fig.~\ref{evn});
(3) PA of the restoring beam (degree);
(4) contour levels (mJy beam$^{-1}$); 
(5) rms level ($\mu$Jy beam$^{-1}$).
\label{contours}}
\begin{tabular}{lcclc}
\toprule
\textbf{Name}	& \textbf{beam} & \textbf{PA} & \textbf{contour levels}	& \textbf{rms}  \\
\midrule
 \multicolumn{5}{c}{{\it e}MERLIN} \\
J2336+0004 &  0.04$\times$0.04   &  0.0   &   0.16$\times$(-1,1,2,3,3.9) &  67 \\
J2346+0059 &  0.04$\times$0.04   &  0.0   &   0.36$\times$(-1,1,2,4,6) &  80 \\
J2357-0010 &  0.04$\times$0.04   &  0.0   &   0.090$\times$(-1,1,2,4)  & 38  \\
J0020-0028 &  0.04$\times$0.04   &  0.0   &   0.12$\times$(-1,1,2,4,5) & 40 \\
J0101-0024 &  0.04$\times$0.04   &  0.0   &   0.13$\times$(-1,1,2,3)   & 61 \\
\midrule
 \multicolumn{5}{c}{JVLA-{\it e}MERLIN} \\
J2336+0004 &  0.06$\times$0.03   & -3.4    &   0.21$\times$(-1,1,1.5,2,3,4.5) &  133 \\
J2346+0059 &  0.08$\times$0.05   & -68.0 &   0.20$\times$(-1,1,2,4,8,16,32) &  52 \\
J2357-0010 &  0.05$\times$0.05   & 0.0    &   0.080$\times$(-1,1,2,4,6,8,10)  & 27  \\
J0020-0028 &  0.10$\times$0.05   & 15.0   &   0.080$\times$(-1,1,2,4,8) & 31 \\
J0101-0024 & 0.08$\times$0.05   &  18.0   &   0.059$\times$(-1,1,2,3,4,6,12)   & 21 \\
\midrule
 \multicolumn{5}{c}{EVN} \\
J0907+3257  &   0.020$\times$0.012 & -54.0  &   0.23$\times$(-1,1,2,4,8,16,32) &  61 \\ 
J0930+3413  & 0.016$\times$0.009    &  -8.1   &   0.24$\times$(-1,1,2,4,8,16) &   33 \\
J0943+3614  &  0.016$\times$0.007 & -23.9  &   2.0$\times$(-1,1,2,4,8,16,32) & 232   \\
J1025+1022  &  0.021$\times$0.011 & -5.4   &   0.98$\times$(-1,1,3,9,27,81) &  139   \\ 
J1040+0910  &  0.014$\times$0.013   &  -28.0   &   0.39$\times$(-1,1,2,3,4,5,5.8) &  102   \\
J1213+5044  &   0.025$\times$0.025   &  0.0  &   2.2$\times$(-1,1,2,4,8,16) &  419   \\
J1230+4700  &  0.024$\times$0.015 & -16.3 & 0.50$\times$(-1,1,2,4,8,16,32,64) &  164  \\
J1530+2705  &  0.021$\times$0.012 &  -10.1 &   1.5$\times$(-1,1,2,4,8,16,32) &  404 \\
J1628+2529  &  0.020$\times$0.012 &  -40.3  &   0.60$\times$(-1,1,3,9,27) & 131   \\
J1703+2410  &  0.015$\times$0.010 &  -14.0 &   0.35$\times$(-1,1,2,4,8,16) & 61   \\
\bottomrule
\end{tabular}
\end{specialtable}


\end{paracol}
\reftitle{References}

\end{document}